\documentclass{article}
\usepackage{PRIMEarxiv}

\usepackage[utf8]{inputenc} % allow utf-8 input
\usepackage[T1]{fontenc}    % use 8-bit T1 fonts
\usepackage{url}            % simple URL typesetting
\usepackage{booktabs}       % professional-quality tables
\usepackage{amsfonts}       % blackboard math symbols
\usepackage{nicefrac}       % compact symbols for 1/2, etc.
\usepackage{microtype}      % microtypography
\usepackage{lipsum}
\usepackage{fancyhdr}       % header
\usepackage{graphicx}       % graphics
\graphicspath{{media/}}     % organize your images and other figures under media/ folder

\def\ci{\perp\!\!\!\perp}

\usepackage{centernot}
\newcommand{\nci}{\centernot{\ci}}

\usepackage[normalem]{ulem}
\newcommand{\stkout}[1]{\ifmmode\text{\sout{\ensuremath{#1}}}\else\sout{#1}\fi}
\newtheorem{thm}{Theorem}
\newtheorem{prop}{Proposition}

\newtheorem{lem}{Lemma}
\newtheorem{ass}{Assumption}

\newenvironment{thma}[1]{\par\noindent{\bf Theorem #1\ }\em}{\em}

\newenvironment{propa}[1]{\par\noindent{\textbf{Proposition #1} }\em}{\em}

\usepackage{color}
\usepackage{xcolor}

\usepackage{subcaption}

\usepackage{amssymb,amsmath}
\allowdisplaybreaks
\usepackage{tikz,pgf}
\usepackage[colorlinks,
            linkcolor=blue,
            anchorcolor=blue,
            citecolor=blue
            ]{hyperref}

\usetikzlibrary{shapes.arrows,shapes}
\usepackage[round]{natbib}

\title{Identification and Estimation for Nonignorable Missing Data:\\ A Data Fusion Approach}
\author{
  Zixiao Wang \\
  Department of Biostatistics \\
  Johns Hopkins University \\
  Baltimore, MD\\
  \texttt{zwang383@jh.edu} \\
   \And
  AmirEmad Ghassami \\
Department of Mathematics and Statistics\\
Boston University\\
Boston, MA\\
  \texttt{ghassami@bu.edu} \\
  \And
  Ilya Shpitser \\
  Department of Computer Science \\
  Johns Hopkins University \\
  Baltimore, MD\\
  \texttt{ilyas@cs.jhu.edu}
}

\begin{document}
\maketitle

\begin{abstract}
We consider the task of identifying and estimating a parameter of interest in settings where data is missing not at random (MNAR). In general, such parameters are not identified without strong assumptions on the missing data model. In this paper, we take an alternative approach and introduce a method inspired by data fusion, where information in an MNAR dataset is augmented by information in an auxiliary dataset subject to missingness at random (MAR).  We show that even if the parameter of interest cannot be identified given either dataset alone, it can be identified given pooled data, under two complementary sets of assumptions.  We derive an inverse probability weighted (IPW) estimator for identified parameters, and evaluate the performance of our estimation strategies via simulation studies, and a data application.
\end{abstract}


\section{Introduction}
%\vspace{-3mm}
Missing data is a pervasive and challenging issue in various applications of statistical inference, such as healthcare, economics, and the social sciences. 
Data are said to be Missing at Random (MAR) when the mechanism of missingness depends only on the observed data. Strategies to deal with MAR have been extensively investigated in the literature \citep{dempster1977maximum, robins1994estimation,tsiatis2006semiparametric, little2019statistical}.
In many practical settings, MAR is not a realistic assumption.  Instead, missingness often depends on variables that are themselves missing. Such settings are said to exhibit nonignorable missingness, with the resulting data being Missing Not at Random (MNAR) \citep{fielding2008simple, schafer2002missing}, 
A classic example of a scenario with MNAR data occurs in longitudinal studies, due to the treatment's toxicity, some patients may become too ill to visit the clinic, leading to the situation where the outcome of certain patients with circumstances associated with those outcomes are more likely to be lost to follow-up \citep{ibrahim2012missing}.

Previous MNAR models typically imposed constraints on target distribution and its missingness mechanism, ensuring the parameter of interest can be identified. This approach goes back to the work of \cite{heckman1979sample}, who proposed an outcome-selection model based on parametric modeling of outcome variable and missing pattern. \cite{little1993pattern} introduced the pattern-mixture model where one needs to specify the distribution for each missing data pattern independently. Other related work involves permutation model \citep{robins1997non}, the discrete choice model \citep{tchetgen2018discrete}, the block-sequential MAR model  \citep{zhou2010block}, 
%item-wise conditionally independent nonresponse (ICIN) 
the no self-censoring (NSC) model \citep{shpitser2016consistent,sadinle2017itemwise,malinsky2021semiparametric}, the instrumental variable approaches \citep{miao2015identification,tchetgen2017general}, and approaches based on graphical models \citep{mohan2013graphical,bhattacharya20completeness} just to name a few.

In some applications where MNAR data is present, researchers may have access to additional auxiliary data on informative variables that are themselves not missing, or missing given a simpler missingness process. Such a \emph{data fusion} strategy is natural in many applications: for example, surveys of HIV patients containing sensitive questions (such as those on sexual history) with high degree of missingness may be augmented with other sources of information with simpler missingness mechanisms, such as electronic health records.

In this paper, we demonstrate that a parameter of interest may be identified given MNAR data in a primary domain, and MAR data from an auxiliary domain, even though the same parameter is not identified from data from either domain alone.
We illustrate our method by
estimating the hospitalization rate in New York during the initial shortage stage of the COVID-19 pandemic in March 2020 where hospitalization status was often unrecorded with a complex censoring mechanism, with the auxiliary data from March 2023, a period characterized by improved conditions and thus reduced level of missingness with a simpler censoring mechanism.

Similar data fusion approaches exist in the literature of causal inference %, where %\ilya{recent work has shown that causal effects that are not identified in the presence of unobserved confounding may become identified given the presence of an auxiliary dataset}
considering settings where the presence of unobserved confounders in the system may render the causal effect of a treatment on an outcome variable unidentified. For instance, works such as
\citep{athey2016estimating,athey2020combining,ghassami2022combining, imbens2022long}
demonstrated that while information is often missing for long-term effects in randomized trials \citep{bouguen2019using}, auxiliary observational studies may contain information on long-term effects of treatments that can render the causal effect identified. 
Note that this is fundamentally different from the case that the parameter is identified in at least one of the domains and the purpose of combining datasets is improvement in estimation efficiency, as opposed to identification (see, for example, \cite{kallus2020role}). 
To the best of our knowledge, our work is the first to pursue data fusion for the purpose of identification in an MNAR setting.

We introduce a novel data fusion technique that combines information from two datasets.
In the first dataset, information on our target variable of interest as well as other variables in the model is only partially observed with an MNAR mechanism, which is more commonly encountered in real-world data. 
In the second data set, the target variable of interest is fully unobserved, and the rest of the variables exhibit missingness with a MAR mechanism.

Using the language of graphical models, we propose two complementary models of the missing mechanisms in the MNAR domain that allow identification of the target parameter to be obtained given data from both domains.
In Model 1, we consider the case where the missingness of the outcome is directly associated with another variable that may itself be potentially missing.  In Model 2, we consider the case where the missingness is directly associated with a potentially missing outcome itself.

The rest of the paper is organized as follows. In Section~\ref{sec:PRELIMINARIES}, we introduce the concepts of missing data and Directed Acyclic Graphs (DAGs), providing an overview of the problem setup and the parameters of interest. 
In Section~\ref{sec:identification}, we describe the assumption of our data fusion setting and present an identification approach from the pooled data under two complementary sets of assumptions. Graphical models are used to illustrate the idea of data fusion and missing mechanisms. 
For estimation, we propose an inverse probability weighted (IPW) estimator for the target parameter in both models, as discussed in section~\ref{sec:estimation}. 
The study of the performance of the proposed estimators and a comparison with a MAR estimator via a series of simulations is presented in Section~\ref{sec:simulation}.
We describe the COVID-19 application analysis in Section~\ref{sec:application}.
We conclude and discuss future work in Section~\ref{sec:discussion}. 
All the proofs are provided in the Appendix.
%\vspace{-3mm}

\section{Preliminaries}
%\vspace{-2mm}
\label{sec:PRELIMINARIES}
\paragraph{Missing Data.} We consider a missing data model which is a collection of distributions that are defined over a set of random variables $\{X, R, Y^{(1)}, Y, M^{(1)}, M\}$. In this context, $X$ represents a set of covariates that are always observed, $Y^{(1)}$ and $M^{(1)}$ represent the underlying outcome variable of interest and another covariate (or set of covariates), respectively, that could be potentially missing, $R$ is a set of binary indicator variables of missingness for the variables $Y^{(1)}$ and $M^{(1)}$, and $Y$ and $M$ represent the observed version of $Y^{(1)}$ and $M^{(1)}$: when $R_Y =R_M= 1$, the corresponding observed variables are $Y \equiv Y^{(1)}$ and $M \equiv M^{(1)}$, when $R_Y =R_M= 0$, $Y = \text{``?''}$ and $M = \text{``?''}$. The full distribution (law) of a missing data model is $p(X,Y^{(1)},M^{(1)},R)$, and it is generally partitioned into two pieces: the target distribution $p(X,Y^{(1)},M^{(1)})$ and the missingness mechanism $p(R|X, Y^{(1)}, M^{(1)})$. While the target distribution consists of potentially missing random variables, the missingness mechanism denotes the patterns exhibited by missingness indicators given the observed and missing variables. The observed distribution is $p(X,R,Y,M)$. When dealing with missing data problems, the objective is to obtain estimations or inferences about functions of variables in $\{X,Y^{(1)},M^{(1)}\}$ based on the observed data. 

\paragraph{Directed Acyclic Graphs (DAGs).} Several widely used missing data models \citep{robins1997non, shpitser2016consistent,miao2015identification,zhou2010block} can be thought of as a factorization of the complete data distribution and represented by a DAG \citep{mohan2013graphical}. A DAG, often denoted as $\mathcal{G}(V)$, is a graph with a vertex set $V$ connected by direct edges, ensuring that no cycles exist within the structure. Statistical models associated with a DAG $\mathcal{G}$ entail probability distributions that factorize as $p(V)=\prod_{V_i \in V} p\left(V_i \mid \mathrm{pa}_{\mathcal{G}}(V_i)\right)$, where $\mathrm{pa}_{\mathcal{G}}(V_i)$ are the parents of node $V_i$ within the DAG $\mathcal{G}$. Conditional independences in any distribution obeying the above factorization may be read off via the d-separation criterion \citep{pearl88probabilistic}. 
Graphical models that allow for context-specific dependence structures have been considered in \citep{nyman2014stratified} and will be employed in our work.

\paragraph{Problem setup.} We consider a setting with two available datasets, a primary dataset and an auxiliary dataset.  The primary dataset is drawn from Domain 1, where the outcome whose mean we would like to estimate is observed but potentially missing not at random.  The auxiliary dataset is drawn from Domain 2, where the outcome is not recorded, but an auxiliary variable set is recorded, but potentially missing at random.

Let $G$ be the binary indicator of the domain: $G=1$ indicates data in the primary domain which is MNAR, and $G=2$ indicates data in the auxiliary MAR domain.  Let $R_1, R_2$ denote indicators for missingness in Domain 1 and Domain 2, respectively. Therefore, observed variables in the second Domain are \{$G=2, X, M, R_2$\}. In the first domain, we can observe \{$G=1, X, M, R_1, Y$\}, where the missing indicator $R_1$ is subject to potentially missing variables $\{M^{(1)},Y^{(1)}\}$,  indicating that $M, Y$ are missing not at random. $R_1$ is the missing indicator for both $M^{(1)}$ and $Y^{(1)}$, i.e., if $R_1=0$, $\{M^{(1)},Y^{(1)}\}$ will be missing at the same time, and therefore $M=``?"$, $Y=``?"$. 
The cause of missingness in Domain 1 can be either $M^{(1)}$, or the outcome itself $Y^{(1)}$; we discussed the identification and estimation of the two cases in Model 1 and Model 2 respectively.
In the pooled dataset containing the collection of random variables $\{G, X, M, R, Y\}$, we define $R=\{R_1,R_2\}$, where $R=R_1$ if $G=1$ and $R=R_2$ if $G=2$.

\paragraph{Estimands.} %Our goal is to infer about the potentially missing outcome mean of $Y^{(1)}$ in the first Domain where $G=1$, 
Our target of inference is the mean of a (potentially missing) outcome $Y^{(1)}$ in the primary domain (denoted by $G=1$).
That is, the parameter of interest in this work is
\[
\beta = \mathbb{E} [Y^{(1)}\mid G =1].
\]
As mentioned earlier, the outcome variable is exclusively present in the primary domain (Domain 1) where we have identification challenges due to MNAR condition. In order to overcome this obstacle, we utilize information from the auxiliary domain (Domain 2) to construct a framework for identification and estimation.
%\vspace{-3mm}
\section{Identification}
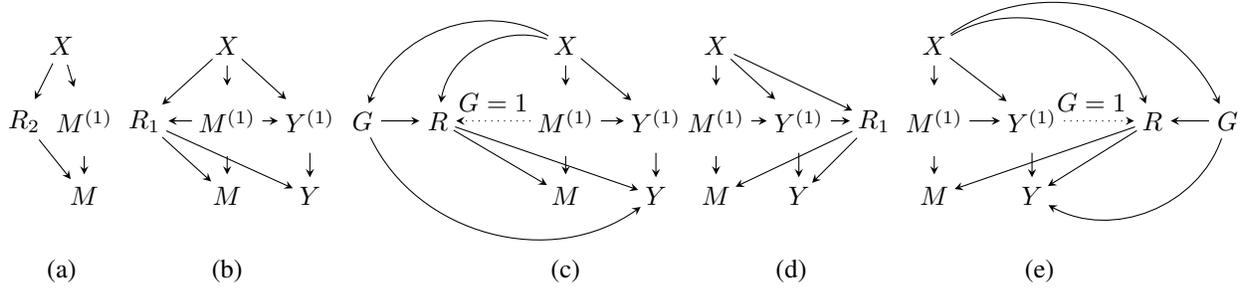
\begin{figure*}[!ht]
\begin{center}
\begin{tikzpicture}[>=stealth, node distance=1cm]
\tikzstyle{format} = [draw=none, thick, circle, minimum size=2.5mm, inner sep=1pt]
\begin{scope}[xshift=-6.6cm]
			\path[->,  thick]
			node[format] (x) {$X$}
			node[format, below of=x, xshift=-0.5cm] (r1) {$R_2$}
			node[format, below of=x, xshift=0.3cm] (m1){$M^{(1)}$}
                node[format, below of=m1] (m) {$M$};
                \node[format, below of=x, yshift=-2.0cm](a){(a)};

			\draw[black,->] (x) edge (r1);
			\draw[black,->] (x) edge (m1);
                \draw[black,->] (r1) edge (m);
                \draw[black,->] (m1) edge (m);             
		\end{scope} 
\begin{scope}[xshift=-4.4cm]
			\path[->,  thick]
			node[format] (x) {$X$}
			node[format, below of=x] (m1) {$M^{(1)}$}
                node[format, left of=m1,xshift=-0.1cm] (r2) {$R_1$}    
			node[format, right of=m1,xshift=0.1cm] (y1) {$Y^{(1)}$}
                node[format, below of=m1] (m) {$M$}
                node[format, below of=y1] (y) {$Y$};
                \node[format, below of=x, yshift=-2.0cm](b){(b)};

			\draw[black,->] (x) edge (m1);
                \draw[black,->] (x) edge (r2);
                \draw[black,->] (x) edge (y1);
                \draw[black,->] (m1) edge (r2);
                \draw[black,->] (m1) edge (y1);
                \draw[black,->] (m1) edge (m);
                \draw[black,->] (y1) edge (y);
                \draw[black,->] (r2) edge (m);
                \draw[black,->] (r2) edge (y);
		\end{scope}
\begin{scope}[xshift=0.1cm]
			\path[->,  thick]
			node[format] (x) {$X$}
			node[format, below of=x] (m1) {$M^{(1)}$}
                node[format,left of = m1,xshift=-0.7cm] (r) {$R$}  
			node[format, right of=m1,xshift=0.2cm] (y1) {$Y^{(1)}$}
                node[format, below of=m1] (m) {$M$}
                node[format, below of=y1] (y) {$Y$}
                node[format, left of = r] (g) {$G$};
                \node[format, below of=x, yshift=-2.0cm](c){(c)};
     
			\draw[black,->] (x) edge (m1);
                \draw[black,->,bend right =50] (x) edge (r);
                \draw[black,->] (x) edge (y1);
                \draw[black,->,dotted] (m1) --node[above]{$G=1$} (r);
                \draw[black,->] (m1) edge (y1);
                \draw[black,->] (m1) edge (m);
                \draw[black,->] (y1) edge (y);
                \draw[black,->] (r) edge (m);
                \draw[black,->] (r) edge (y);
                \draw[black,->] (g) edge (r);
                \draw[black,->, bend right =50] (x) edge (g);
                \draw[black,->,bend right =50] (g) edge (y);
				\end{scope}	
\begin{scope}[xshift=2.1cm]
	\path[->,  thick]
			
	node[format] (x) {$X$}
	node[format, below of=x] (m1) {$M^{(1)}$}       
        node[format, right of=m1, xshift=0.1cm] (y1) {$Y^{(1)}$}
        node[format, below of=m1] (m) {$M$}
        node[format, below of=y1] (y) {$Y$}
        node[format, right of=y1] (r2) {$R_1$};
        
        \node[format, below of=x, yshift=-2.0cm, xshift=1.0cm](d){(d)};

	\draw[black,->] (x) edge (m1);
        \draw[black,->] (x) edge (r2);
        \draw[black,->] (x) edge (y1);
        \draw[black,->] (y1) edge (r2);
        \draw[black,->] (m1) edge (y1);
        \draw[black,->] (m1) edge (m);
        \draw[black,->] (y1) edge (y);
        \draw[black,->] (r2) edge (m);
        \draw[black,->] (r2) edge (y);
   
		\end{scope}
\begin{scope}[xshift=5.0cm]
		\path[->,  thick]
		node[format] (x) {$X$}
	node[format, below of=x] (m1) {$M^{(1)}$}
	node[format, right of=m1,xshift=0.3cm] (y1) {$Y^{(1)}$}
        node[format,right of = y1,xshift=0.6cm] (r) {$R$} 
        node[format, below of=m1] (m) {$M$}
        node[format, below of=y1] (y) {$Y$}
        node[format, right of = r] (g) {$G$};
      \node[format, below of=x, yshift=-2.0cm, xshift=1.4cm](e){(e)};

	\draw[black,->] (x) edge (m1);
        \draw[black,->,bend left =50] (x) edge (r);
        \draw[black,->] (x) edge (y1);
        \draw[black,->,dotted] (y1) --node[above,xshift=-0.1cm]{$G=1$} (r);
        \draw[black,->] (m1) edge (y1);
        \draw[black,->] (m1) edge (m);
        \draw[black,->] (y1) edge (y);
        \draw[black,->] (r) edge (m);
        \draw[black,->] (r) edge (y);
        \draw[black,->] (g) edge (r);
        \draw[black,->,bend left =50] (x) edge (g);
        \draw[black,->,bend left =50] (g) edge (y);
		\end{scope}
\end{tikzpicture}
\end{center}
\vspace{-0.5cm}
\caption{Graphical models: (a) The auxiliary MAR data domain in both Model 1 and Model2 .
(b) Primary MNAR domain in Model 1.
(c)The pooled data for Model 1, including the selection at random mechanism. 
(d) Primary MNAR domain in Model 2.
(e) The pooled data for Model 2, including the selection at random mechanism. Notice the text $G=1$ on the dotted arrow denotes a context-dependent relationship.
}%\vspace{-3mm}
\label{fig:dag_model12}
\end{figure*}

%\vspace{-2mm}
\label{sec:identification}
In this section, we examine two complementary sets of assumptions pertaining to the identification of  Model 1 and Model 2. Subsequently, we establish the corresponding identification theorems. Our discussion commences with assumptions shared by both models in Section~\ref{sec:bothassumption}, followed by a separate exploration of additional assumptions and identification for each model. Before proceeding further, we note that our assumptions lead to graphical models of missing data spanning two domains, with the conditional independences defining the model illustrated (via the d-separation criterion) in the graphs shown in Fig.~\ref{fig:dag_model12}.  Note that any graphical model implying the same conditional independence restrictions as these figures also implies our model.
%\vspace{-2mm}
\subsection{Data Fusion Assumptions}
%\vspace{-2mm}
\label{sec:bothassumption}
The full data distribution in the MNAR dataset (Domain 1), denoted as $p(X, M^{(1)}, Y^{(1)}, R_1, G=1)$,
cannot be deduced solely from the observed data distribution $p(X, M, Y, R_1, G=1)$, unless certain constraints are placed on the mechanism responsible for data missingness. In our analysis, we leverage the information in Domain 2. In this domain, data is missing at random which is formalized in the following.

\begin{ass}[Auxiliary Domain MAR] 
\label{ass1:mr_g1} 
In auxiliary domain ($G=2$),  $M^{(1)}$ is missing at random, i.e., the missing mechanism is independent of potentially missing variable $M^{(1)}$ conditional on observed covariates $X$. That is,
    \begin{align}
    M^{(1)} \ci R \mid X,G=2
    \label{equ:ass1}
\end{align}
\end{ass}

{In order for us to be able to leverage the auxiliary domain, the information encoded in this domain must be relevant to the primary domain. Such a relevance requirement is usually stated by a \emph{external validity}-type assumption in the literature \citep{hotz2005predicting}. We require the relevance of the information in the two domains in the form of the following \emph{selection} assumption.}

\begin{ass}[Selection at random] 
\label{ass2:mg_x}
Given covariates $X$, the domain indicator $G$ is conditionally independent of $M^{(1)}$. That is,
 \begin{align}
      M^{(1)} \ci G \mid X
     \label{equ:ass2}
 \end{align}
\end{ass}
Assumption~\ref{ass2:mg_x} limits the differences of $M^{(1)}$ between the two populations by requiring that conditioned on the rest of the covariates, it is as if units are randomly selected to belong to the domains. In practice, this assumption should be discussed and justified by domain experts when combining the datasets.

\subsection{Model 1}
%\vspace{-2mm}
In Model 1, we consider the scenario where the covariates $X$ and the potentially missing variable $M^{(1)}$ contribute to the missingness in the primary domain, yet the outcome variable is not directly involved in the missingness mechanism, i.e., it is independent of the missing indicator conditioned on $M^{(1)}$ and $X$. This restriction is formulatted in the following Assumption~\ref{ass3:yr_xmg2}.

\begin{ass}[Primary Domain MNAR]
\label{ass3:yr_xmg2}
In primary domain ($G=1$), the missing mechanism is independent of the outcome $Y^{(1)}$ given both $M^{(1)}$ and covariates $X$.
\begin{align}
    Y^{(1)} \ci R \mid X, M^{(1)} , G=1.
    \label{equ:ass3}
\end{align}
\end{ass}

Figures~\ref{fig:dag_model12} (a), (b) and (c) show graphical representations for Domain 1, Domain 2 and the pooled dataset, respectively, which satisfy Assumptions \ref{ass1:mr_g1} and \ref{ass3:yr_xmg2}. 
Fig.~\ref{fig:dag_model12} (a) represents the graphical model in the Domain 2 (which is the pooled data set conditioned on $G$ = 2). Notice that variable $Y$ is absent in the second Domain. Fig.~\ref{fig:dag_model12} (b) represents the graphical model in Domain 1 (which is the pooled data set conditioned on $G$ = 1). 
$X$ is the same list of covariates as that in Domain 1, and $R_1$ is the missing indicator for both $M^{(1)}, Y^{(1)}$ in Domain 1. We represent the  graphical model of both domains pooled together in Fig.~\ref{fig:dag_model12} (c), where the text $G=1$ on the dotted arrow from $M^{(1)}$ to $R$ denotes a conditional relationship that $M^{(1)}$ is parent of $R$ if and only if $G=1$. It is important to note that the presented graphical models in Fig.~\ref{fig:dag_model12} are only one example of models that satisfy our assumptions. For instance, the edges between $Y^{(1)}$ and $M^{(1)}$ and $X$ and $G$ can be reversed without changing the model, and thus the identifying assumptions.

\begin{thm}[Identification in Model 1]
\label{thm:identification1}
Under Assumptions~\ref{ass1:mr_g1},~\ref{ass2:mg_x}, and~\ref{ass3:yr_xmg2}, parameter $\beta = \mathbb{E} [Y^{(1)}|G =1]$ is identified using the following functional
%for $\mathbb{E}[Y^{(1)}|G=1]$:
\begin{equation}
\label{equ:identification1}
\begin{aligned}
\mathbb{E}[\mathbb{E}[g_1(X,M)  \mid X,G=2, R=1 ]\mid G=1],
\end{aligned}
\end{equation}
where $g_1(X,M) \equiv \mathbb{E}[Y |X,M,G=1,R=1]
$.
\end{thm}

The %identification procedure
identifying functional involves three distributions: $p(Y \mid X,M,G=1,R=1)$, $p(M \mid X, G=2, R=1)$, and $p(X \mid G=1)$.
Notice that $p(X \mid G=1)$ involves no potentially missing variable and is thus identified. Distributions $p(Y \mid X,M,G=1,R=1)$ and $ p(M \mid X, G=2, R=1)$ 
are also identified as they are conditioned on $R=1$.
As a result, the mean of the potentially missing outcome variable in the first domain can be identified from the observational data.

%\vspace{-2mm}
\subsection{Model 2}
%\vspace{-2mm}
In Model 2, we consider a scenario where the potentially missing outcome $Y^{(1)}$ is directly associated with the missingness indicator in Domain 1. This necessitates assumptions complementary to those in Model 1 for the identification and estimation of the outcome mean.
Inspired by the shadow variable approach in the study of MNAR data \citep{miao2015identification}, we consider the following assumption. 
\begin{ass}
\label{ass4}
[\cite{miao2015identification}] 
The potentially missing variable $M^{(1)}$ satisfies the following conditional independence requirements 
\begin{align}
& M^{(1)} \ci R \mid X, Y^{(1)} , G=1  \\
& M^{(1)} \nci Y^{(1)} \mid X, R=1 , G=1
\end{align}
\end{ass}
Assumption~\ref{ass4} formalizes the idea that the missingness process in Domain 1 may depend on $(X, Y^{(1)})$, but not on the potentially missing variable $M^{(1)}$ after conditioning on $(X, Y^{(1)})$. One notable distinction from the assumption in \citep{miao2015identification} is that the counterpart of variable $M^{(1)}$ in their setting, which is called the shadow variable, is fully observed, whereas our framework encompasses scenarios where this variable may potentially be missing. Furthermore, the outcome variable serves as both the cause of its own missingness and the missingness of the variable $M^{(1)}$. This, in turn, creates a situation characterized by non-random missingness.

The graphical model for Model 2 is shown in Fig.~\ref{fig:dag_model12} (a), (d) and (e). The structure of Domain 2 of Model 2 (Fig.~\ref{fig:dag_model12} (a)) is the same as that in Model 1. However, for Domain 1, a key difference  is that $Y^{(1)}$ contributes to the missingness instead of $M^{(1)}$, as is shown in Figure~\ref{fig:dag_model12} (d), while Figure~\ref{fig:dag_model12} (e) shows both domains pooled together.

As in the shadow variable approach of \cite{miao2015identification}, we leverage the odds ratio function to encode the deviation between the observed and missing data distributions in Domain 1, which is defined as
\begin{align}
&OR\left(Y^{(1)}, X, M^{(1)}\right) =\frac{p\left(Y^{(1)} \mid R=0, X, M^{(1)},G=1\right)}{p\left(Y^{(1)} \mid R=1, X, M^{(1)},G=1\right)} \times  \frac{p\left(Y^{(1)}=0 \mid R=1, X, M^{(1)},G=1\right)}
{p\left(Y^{(1)}=0 \mid R=0, X, M^{(1)},G=1\right)}.
\end{align}

The following proposition shows that a number of parts of the full data distribution are identified under the assumptions in Model 2.  These results will allow us to obtain identification of the target parameter.
\begin{prop}[\cite{miao2015identification}]
\label{prop:model2}
Under Assumption~\ref{ass4}, for all $(X, Y^{(1)}, M^{(1)})$ in Domain 1, we have
\begin{align}
OR\left(Y^{(1)}, X, M^{(1)}\right)
&=OR\left( Y^{(1)},X\right)\notag \\ 
&= \frac{p\left(R=0 \mid X, Y^{(1)}, G=1\right)}{p\left(R=1 \mid X, Y^{(1)}, G=1 \right) } \times \frac{ p\left(R=1 \mid X, Y^{(1)}=0, G=1\right)}{p\left(R=0 \mid X, Y^{(1)}=0,G=1\right)} 
\label{equ:or}
\end{align}

\begin{align}
p\left(Y^{(1)} \mid R=0, X, M^{(1)}, G=1\right) = \frac{p\left(Y^{(1)} \mid R=1, X, M^{(1)}, G=1\right) OR\left(X, Y^{(1)}\right)}{\mathbb{E}\left[OR\left(X, Y^{(1)}\right) \mid R=1,X,M^{(1)},G=1\right]}
\label{equ:y1_r0_x1}
\end{align}
%\vspace{-0.3cm}
\begin{align}
&p\left(R=1 \mid X, Y^{(1)}, G=1\right)^{-1} = 1 + \frac{OR\left(X, Y^{(1)}\right) p\left(R=0 \mid X, Y^{(1)}=0, G=1\right)}{p\left(R=1 \mid X, Y^{(1)}=0, G=1\right)}
\label{equ:model2_r1_xy1g2}
\end{align}
\begin{align}
&p\left(R=1 \mid X, Y^{(1)}=0, G=1\right) =\frac{\mathbb{E}\left[OR\left(X, Y^{(1)}\right) \mid R=1,X,G=1\right]}{\mathbb{E}\left[OR\left(X, Y^{(1)}\right) \mid R=1,X,G=1\right] + \frac{p(R=0 \mid X, G=1)}{p(R=1 \mid X,G=1)}}
\label{equ:r1_x_y0_g2}
\end{align}
\begin{align}
\label{equ:ortilde}
&\mathbb{E}\{\widetilde{OR}\left(X, Y^{(1)}\right)  \mid R=1,X,M^{(1)},G=1\} = \frac{p\left(M^{(1)} \mid X, R=0, G=1\right)}{p\left(M^{(1)} \mid X, R=1, G=1\right)}
\end{align}
where,
\begin{align}
\label{equ:ortildeor}
 \widetilde{OR}\left(X, Y^{(1)}\right) = \frac{OR\left(X, Y^{(1)}\right)}{\mathbb{E}\{OR\left(X, Y^{(1)}\right) \mid R=1,X,G=1\}} 
\end{align}
\end{prop}

%These results only rely on Assumption \ref{ass4}.
%In these results, we use the value 0 as the \emph{baseline value} for the variables.
%For completeness,
%\vspace{-2mm}
We present the proof of these results in the Appendix. 
Equation (\ref{equ:or}) shows that the odds ratio function in the MNAR domain is only related to $\{X, Y^{(1)}\}$ under Assumption~\ref{ass4}. 
Equation (\ref{equ:y1_r0_x1}) shows that the missing data distribution of the outcome can be recovered by imposing odds ratio function and the complete case distribution. 
Equation (\ref{equ:model2_r1_xy1g2}) shows that the propensity of missingness given the outcome, $p\left(R=1 \mid X, Y^{(1)}, G=1\right)$, can be recovered by the odds ratio function and baseline propensity score $p\left(R=1 \mid X, Y^{(1)}=0, G=1\right)$, while the baseline propensity score depend on the odds ratio function with $ p (R=1 | X, G=1)$ obtained as stated in Equation (\ref{equ:r1_x_y0_g2}).

In light of Proposition~\ref{prop:model2}, it becomes evident that the crux of the matter lies in the identification of the odds ratio function. Equation (\ref{equ:ortilde}) serves as a pivotal mathematical expression for $OR(X,Y^{(1)})$. With $p(M^{(1)}|X,R=0,G=1), p (M^{(1)} |X,R = 1, G=1)$ and $p (Y^{(1)} | M^{(1)},X,R = 1,  )$, equation (\ref{equ:ortilde}) is a Fredholm integral equation of the first kind with $\widetilde{OR}(X, Y^{(1)} )$ to be solved for. However, the distribution of $p\left(M^{(1)} \mid X, R=0, G=1\right)$ cannot be observed directly from Domain 1. 

Therefore, to identify the missing distribution $p\left(M^{(1)} \mid X, R=0, G=1\right)$ effectively, we employ the observed distribution in Domain 2 by noticing that, under Assumption~\ref{ass1:mr_g1} and~\ref{ass2:mg_x}, we have 
\begin{equation}
\label{equ:fm_x_r0_g2}
\begin{aligned}
      p(M^{(1)}|X, R=0, G=1)p(R=0|X, G=1)
      %&=p(M^{(1)}|X, G=1)-p(M^{(1)}|X, R=1, G=1)p(R=1|X, G=1)\\
      %&=p(M^{(1)}|X, G=2)-p(M|X, R=1, G=1)p(R=1|X, G=1)\\
      %&=p(M^{(1)}|X,R=1, G=2)-p(M|X, R=1, G=1)p(R=1|X, G=1)\\
      =p(M|X,R=1, G=2) -p(M,R=1,|X, G=1).
\end{aligned}
\end{equation}
As the distributions on the right-hand side are all observed, we conclude that the odds ratio $OR(X, Y^{(1)})$ is identified, and hence, $\widetilde{OR}(X, Y^{(1)})$ is identified by the formula for $\widetilde{OR}(X, Y^{(1)})$ in Equation (\ref{equ:ortildeor}). Note that equation (\ref{equ:fm_x_r0_g2}) serves as a bridge, allowing us to leverage information from domain 2 to recover the missing distribution in domain 1.

To guarantee the uniqueness of the solution for Equation (\ref{equ:ortilde}), we assume the condition below stands.
\begin{ass}{}
Completeness of $p(Y | R = 1, X, M, G=1)$. For all square integrable functions $h(X, Y^{(1)})$, $\mathbb{E}{[h(X,Y^{(1)}) \mid R = 1, X, M,G=1}] = 0$ almost surely if and only if $h(X, Y^{(1)} ) = 0 $ almost surely.
\label{ass5:completeness}
\end{ass}

\begin{thm}{(Identification in Model 2)}
\label{thm:identification2}
Under Assumptions~\ref{ass1:mr_g1},~\ref{ass2:mg_x},~\ref{ass4} and~\ref{ass5:completeness}, parameter $\beta = \mathbb{E} [Y^{(1)}|G = 1]$ is identified using the following formula:
\begin{align}
\label{equ:identified2}
&\mathbb{E}[Y^{(1)}|G=1]  \\
&= \sum_{y,m,x} y p(Y^{(1)}=y|X=x,R=1,G=1, M^{(1)}=m) \notag \\
&\times \Biggl(
p(M^{(1)}=m,X=x, R=1 | G=1)  \notag\\
&+  \frac{ OR\left(X=x, Y^{(1)}=y\right)}{\mathbb{E}\left[OR\left(X, Y^{(1)}\right) \mid R=1,X=x,M^{(1)}=m,G=1\right]} \notag\\
&\times \Bigr[\frac{p(M^{(1)}=m|X=x,R=1, G=2)}{p(R=0|X=x, G=1)} -\frac{p(M^{(1)}=m,R=1|X=x,  G=1)}{p(R=0|X=x, G=1)}\Bigr]\notag\\
&\times p(X=x,R=0|G=1) 
\Biggl)
\end{align}
\end{thm}

%Assumption~\ref{ass4} results in Equation (\ref{equ:or}) for the odds ratio function, in which the identification of the distribution $p\left(M^{(1)} \mid X, R=0 , G=1\right)$ in Domain 1 is guaranteed by equation (\ref{equ:fm_x_r0_g2}).  After identifying the odds ratio function, one can recoverfrom  Equation (\ref{equ:y1_r0_x1}) and therefore, the third component $p(Y^{(1)}|X,R,G=1, M^{(1)})$ is identified.

\section{Estimation}
%\vspace{-2mm}
\label{sec:estimation}
In this section, we focus on the estimation aspect of the problem and propose estimation strategies for the identified functionals for our target parameter under Model 1 and Model 2. We propose inverse probability weighting (IPW) estimators for both models, which rely on the propensity scores. In what follows, $\hat{\mathbb{E}}$ denotes the empirical mean operator.

%\vspace{-2mm}
\subsection{Model 1}
%\vspace{-2mm}
Let $q(X,M^{(1)})=1/p(R=1 \mid X, M^{(1)}, G=1)$. 
Our estimation strategy for the parameter in Equation (\ref{equ:identification1}) is as follows.

\begin{prop}
\label{prop:model1.ipw}
Under Assumptions~\ref{ass1:mr_g1},~\ref{ass2:mg_x}, and~\ref{ass3:yr_xmg2}, a correctly specified working model for $p(R=1 \mid X, M^{(1)}, G=1; \alpha)$, and the regularity conditions for estimating equations described by \cite{newey1994large}, using a user-specified vector function $h(X, M^{(1)})$, the IPW estimator, denoted by $\beta^{IPW}$, obtained by solving
\begin{align}
 &\mathbb{\hat{E}}\left[ \left( q(X,M^{(1)};\hat{\alpha})R \cdot h(X,M^{(1)})  \right) \mid G=1\right]  \\
&- \mathbb{\hat{E}}\left[  \mathbb{\hat{E}}\left[ h(X,M^{(1)}) \mid X, R=1, G=2 \right] \mid G=1\right]=0 \notag\\
 &\mathbb{\hat{E}}\left[ q(X,M^{(1)};\hat{\alpha})R\cdot Y^{(1)}   - \hat{\beta_{IPW}}|G=1\right]=0 
\end{align}
is a consistent estimator.

\end{prop}

%Notice, although the propensity score $p(R=1 \mid X, M^{(1)}, G=1)$ is not observed, we only use the complete case for estimation.

%\vspace{-2mm}
\subsection{Model 2}
%\vspace{-2mm}
We next consider the estimation of the parameter of interest in Model 2. In the context of the shadow variable configuration, \cite{miao2015identification} formerly introduced an estimator relying on odds ratio and baseline propensity score in MNAR data. We extend that strategy here by leveraging information from Domain 2 to compensate for the fact that variable $M^{(1)}$ might be missing.
%proposing working models for the odds ratio function  $OR(X, Y^{(1)}; \gamma)$ in Domain 1 and the baseline propensity score $p(R = 1 | X, Y^{(1)} = 0,G=1; \alpha)$, an additionally,\stkout{ a working model for $p(M \mid X,R=1,G=2)$ estimated from Domain 1,}} together to recover the propensity score in equation (\ref{equ:model2_r1_xy1g2}). 
Let $w(X, Y^{(1)},G=1; \alpha, \gamma) = 1/p (R = 1 | X, Y^{(1)},G=1; \alpha, \gamma)$.
Our estimation strategy for the parameter in Equation (\ref{equ:identified2}) is as follows.
%, we can solve for $\hat{\alpha}, \hat{\gamma}$ and $\hat{\beta}_{IPW}$ using the system of estimating equations in the following proposition.
%we obtain $\hat{\alpha}, \hat{\gamma}$ and the inverse probability weighted estimator $\hat{\beta_{IPW}}$.
\begin{prop}
\label{prop:model2.estimation}
Under Assumptions~\ref{ass1:mr_g1},~\ref{ass2:mg_x},~\ref{ass4} and ~\ref{ass5:completeness}, a correctly specified working model for odds ratio function $OR(X, Y^{(1)}; \gamma)$ and the baseline propensity score $p(R = 1 | X, Y^{(1)} = 0,G=1; \alpha)$, and the regularity conditions for estimating equations described by \cite{newey1994large}, using a user-specified vector function $h(X, M^{(1)})$, the IPW estimator, denoted by $\beta^{IPW}$, obtained by solving
\begin{align}
 &\mathbb{\hat{E}}\left[ \left( w(X,Y^{(1)};\hat{\alpha},\hat{\gamma})R \cdot h(X,M^{(1)})  \right) \mid G=1\right]  \\
&- \mathbb{\hat{E}}\left[  \mathbb{\hat{E}}\left[ h(X,M^{(1)}) \mid X, R=1, G=2 \right] \mid G=1\right]=0 \notag \\
&\mathbb{\hat{E}}\left[ (w(X,Y^{(1)};\hat{\alpha},\hat{\gamma})R\cdot Y^{(1)}   - \hat{\beta}_{IPW}|G=1\right]=0 
\end{align}
is a consistent estimator.
\end{prop}
%\vspace{-3mm}
\section{Simulation Study}
%\vspace{-2mm}
\label{sec:simulation}
\begin{figure*}[h]
  \begin{subfigure}{0.5\textwidth}
    \centering
\includegraphics[width=0.95\linewidth]{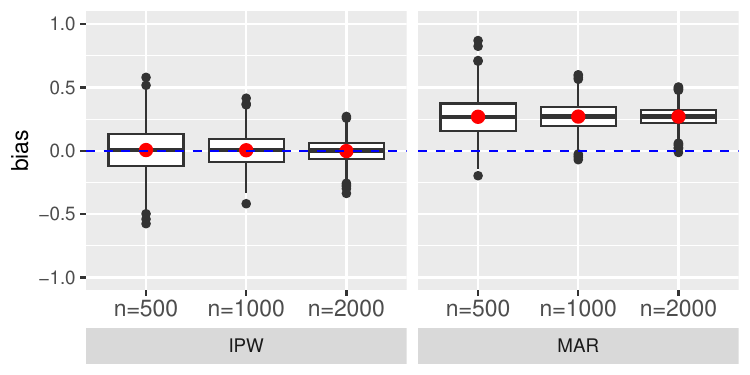}
    \caption{Model1 - T: model is correctly specified.}
    \label{fig:model1_T}
  \end{subfigure}%
  \begin{subfigure}{0.5\textwidth}
    \centering
    \includegraphics[width=0.95\linewidth]{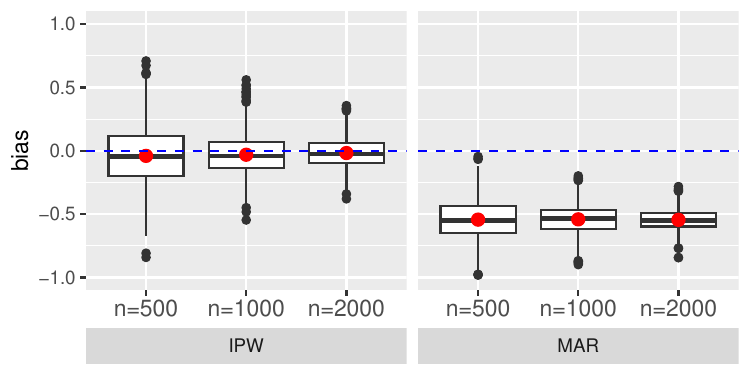}
    \caption{Model1 - F: model is misspecified.}
    \label{fig:model1_F}
  \end{subfigure}

  \begin{subfigure}{0.5\textwidth}
    \centering
    \includegraphics[width=0.95\linewidth]{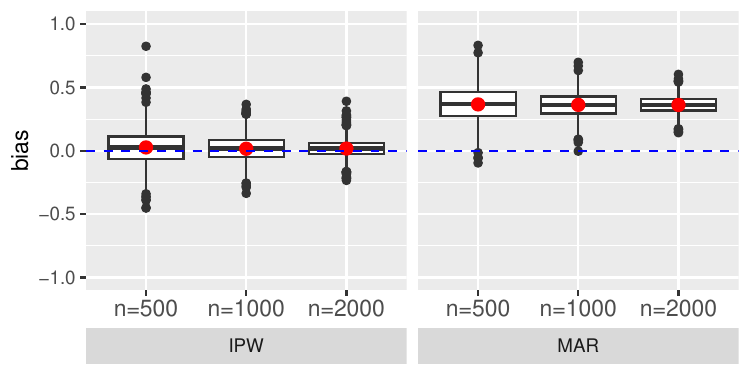}
    \caption{Model2 - T: model is correctly specified.}
    \label{fig:model2_T}
  \end{subfigure}%
  \begin{subfigure}{0.5\textwidth}
    \centering
    \includegraphics[width=0.95\linewidth]{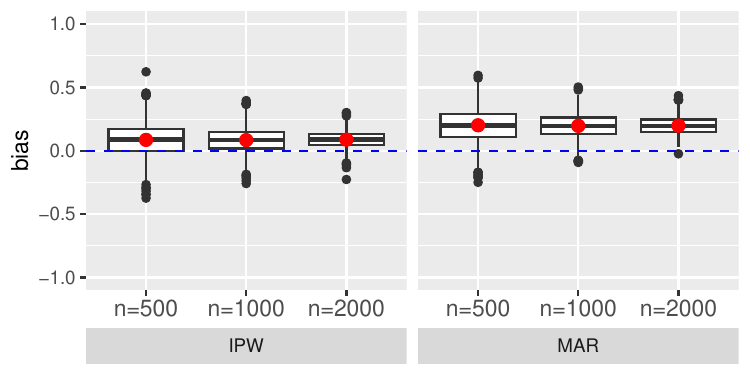}
    \caption{Model2 - F: model is misspecified.}
    \label{fig:model2_F}
  \end{subfigure}
  \caption{Simulation results for Model 1 and Model 2:  Bias for estimation of $\mathbb{E} [ Y^{(1)}\mid G=1 ]$. Boxplots of correct and misspecified settings, calculated from 1000 trials at sample sizes $n \in \{500, 1000,2000\}$. The red point indicates the mean. Statistics of boxplots are in Table~\ref{tab:box.m1} and Table~\ref{tab:box.m2}.
  (a) MAR estimates are clearly biased upwards. (b) IPW estimates, though slightly biased, concentrate around the true value as sample size increases. (c) IPW estimates are less biased than MAR. (d) IPW estimates are less biased than MAR estimates. Statistics of boxplots are in Table~\ref{tab:box.m1} and Table~\ref{tab:box.m2} in Appendix.}%\vspace{-3mm}
  \label{fig:model12_boxplot}
\end{figure*}

We investigate the performance of the proposed framework in Section~\ref{sec:estimation} to estimate the outcome means in the primary domain, $\beta= \mathbb{E}[Y^{(1)} \mid G=1]$, through a comprehensive series of simulation experiments. For each model, we generate a data consistent with the correctly specified working model in estimation (T), and a misspecified one to check robustness (F).
In both models, to streamline the estimation process, we notice that flipping the edge from $X$ to $G$ does not alter the underlying graph structure in Fig.~\ref{fig:dag_model12}. Consequently, we opt to employ $G$ as the primary basis for estimation.
%\vspace{-2mm}
\subsection{Simulation of Model 1}
%\vspace{-2mm}
We first generate a binary grouping variable $G$ with $n$ observations. Each observation in $G$ is randomly assigned a value of $1$ or $2$ with equal probability, dividing the dataset into Domain 1 ($G=1$) and Domain 2 ($G=2$).

For the auxiliary domain ($G=2$), we assume the following models for covariates $X$, variable $M^{(1)}$, and the propensity score:
\begin{align}
  &X \mid G=2 \sim \mathcal{N}\left(0,1\right) \notag\\
  &M^{(1)} \mid X, G=2  \sim 
  \mathcal{N}\left(0.4X^2,1\right) \notag \\
&p(R=1\mid X,G=2)= logis(1.4+X) \notag
\end{align}
where $logis(x)=(1 + \exp(-x))^{-1}$. Under this setting, the missing data proportion of $M$ in Domain 2 is between 50\% and 60\%.

For primary domain ($G=1$), we posit varying distributions for the covariates $X$, while maintaining the same distribution for $M^{(1)}$ given $X$, satisfying Assumption \ref{ass2:mg_x}. Moreover, we assume the distributions for $Y^{(1)}$ is a function of both $X$ and $M^{(1)}$ by Assumption~\ref{ass3:yr_xmg2}:
\begin{align}
&X \mid G=1 \sim \mathcal{N}\left(1,1\right) \notag\\
&M^{(1)} \mid X, G=1  \sim \mathcal{N}\left(0.4X^2,1\right) \notag \\
&Y^{(1)} \mid X, M^{(1)}, G=1  \sim \mathcal{N}\left(X + M^{(1)},1\right) \notag \\
&p(R=1\mid X,M^{(1)},G=1)= logis(0.3 + 0.1X + M^{(1)}) \notag
\end{align}
Thus, the missing data proportion is between 20\% and 40\%.
%\vspace{-2mm}

For our IPW estimation approach, we specify a working model for $p(R=1\mid X,M^{(1)},G=1)=\alpha_0 + \alpha_1 X+\alpha_2 M^{(1)}$, and a set of correctly specified estimation equations for $\mathbb{E}[M^{(1)} \mid X, R=1, G=2]$. Considering total sample sizes $n \in\{ 500, 1000, 2000\}$, we summarize the results using boxplot in Fig.~\ref{fig:model1_T}. We generate additional misspecified data by replacing with $p(R=1\mid X,M^{(1)},G=1)= logis(0.3 + 0.1X - [M^{(1)}]^2) $, and summarized in Fig.~\ref{fig:model1_F}. 

For each scenario, we compare the result of our IPW estimator with a naive estimator assuming Missing at Random (MAR), which is derived through linear regression for $Y$ given $X$ on the complete cases. 

For a specific pooled dataset, we used bootstrapping of size $k=1000$ to report correctly specified IPW estimation result and 95\% confidence interval (95\% CI), as shown in Table~\ref{tab:boot_m1}.

{\scriptsize
\begin{table}[h]
\caption{Bootstrap confidence intervals for Model 1 (T)\\ 
(True value of $\beta=1.8$).} 
\label{tab:boot_m1}
\begin{center}
\begin{tabular}{l|lllll}
\hline
\textbf{n}  &\textbf{Est.} & \textbf{95\% CI} &    \textbf{Width} &\textbf{Bias} \\
\hline %\\
$500$    &1.941 & [1.594, 2.275] & 0.681 & 0.141 \\
$1000$   &1.744 & [1.508, 1.970] & 0.462 & -0.056 \\
$2000$    &1.767 & [1.598, 1.930] & 0.332 & -0.033 \\
\hline
\end{tabular}
\end{center}
\end{table}

}
{\scriptsize
\begin{table}[h]
\caption{Bootstrap confidence intervals for Model 2 (T).\\
(True value of $\beta=-0.659$).} 
\label{tab:boot_m2}
\begin{center}
\begin{tabular}{l|lllll}
\hline
\textbf{n}  &\textbf{Est.} & \textbf{95\% CI} &    \textbf{Width} & \textbf{Bias} \\
\hline %\\
500 &  -0.557 &[-0.815, -0.303]& 0.513 & 0.102 \\
n=1000& -0.760 &[-0.949, -0.569]& 0.380 & -0.101 \\
n=2000& -0.707 &[-0.841, -0.561]& 0.281 & -0.048 \\
\hline
\end{tabular}
\end{center}
\end{table}
}
\subsection{Simulation of Model 2}
%\vspace{-2mm}
The data generating process for Model 2 is slightly different due to the factorization.  We started by generating the binary grouping variable $G$ as that for Model 1. For primary domain ($G=1$), we generate a covariate $X | G=1 \sim N(0,1)$, and then generate $(Y,M,R)$  as shown below:
\begin{align}
&M \mid R=1, X, G=1  \sim \mathcal{N}\left(-0.4X^2, 1\right)\notag\\
&Y \mid R=1, X, M, G=1 \sim \mathcal{N}\left(X+ M, 1\right)\notag\\
&\operatorname{logit} p(R=1 \mid X, Y^{(1)}=0, G=1)  =0.5 + 0.4 X\notag \\
&\operatorname{OR}(X, Y^{(1)}; G=1) =\exp (-0.3 Y^{(1)}) \notag
\end{align}
For the auxiliary domain ($G=2$), we followed the same generating process for $\{X,M^{(1)}\}$ as above, but a different missing mechanism, $p(R=1\mid G=2,X) =logis(X)$.

In these specified conditions, the missing rate is between 40\% and 50\% in both domains. We employed correctly specified estimation equations for $\mathbb{E}[M^{(1)} \mid X, R=1, G=2]$, and the above working models in Domain 2. We also generate a misspecified setting (F) by replacing with $  logit [p (R = 1 | Y^{(1)} = 0, X,G=1)] = 0.5 + 0.4X + 0.4X^2$
but always impose first-order linear regression. We simulate $1000$ replicates, with sample sizes $n \in \{500, 1000, 2000\}$. We present the results using boxplots with a comparison of MAR estimator in Fig.~\ref{fig:model2_T} and Fig.~\ref{fig:model2_F}, and summarize the statistics of boxplots in in Appendix Table~\ref{tab:box.m2}. Bootstrapping result is reported in Table~\ref{tab:add2}.

%\vspace{-3mm}
\section{Application to COVID-19 case data}
\label{sec:application}
In this section, we employ both Model 1 and Model 2 to analyze COVID-19 case data in New York State (NYS), focusing on estimating the hospitalization rate during the initial wave of the pandemic around March 2020.

Due to surge of COVID-19 cases, the situation in NYS hospitals in March 2020 was critical, with reported shortages, and a triage system for patient care, leaving some 
% regarded as paramedic shortages as reported in the city's 911 emergency response system, when emergency workers had to decide which cases to prioritize, and some patients were left home
without adequate care \citep{schmitt2020covid, Watkins2020}. By March 2023, COVID-19 was well on the way to endemicity, and additional resources had been committed to COVID-19 patient care.  Both domains exhibit missingness in hospitalization status.  However, in our analysis, we assume March 2020 data to exhibit complex MNAR missingness due to systemic early difficulties with COVID-19 respnose, while later data from March 2023 to exhibit a more manageable MAR missingness.
%In addition,
%the city had undertaken significant efforts and resulted in improvement in the availability of medical resources. Therefore, we consider race and county information collected in March 2023 as an auxillary dataset
%assuming shared race distribution across time.  

\subsection{Data}
The data source is the COVID-19 Case Surveillance Public Use Data with Geography maintained by  \citep{CDC2024}. New York State was selected as the target population, with data in March 2020 as the primary domain and in March 2023 as the auxiliary domain. We applied filtering procedures as detailed in the Appendix \ref{sec:app_covid}, resulting in a sample size of $78119$ in March 2020 and $38237$ in March 2023.

In this dataset, let $X$ represent the patient's county, recognizing that spatial information plays a crucial role in accounting for the spread of infectious diseases, $Y$ be a binary indicator of whether the patient reported as 'Hospitalized' and $M$ be the race of the patient. The missing rate is $70\%$ in March 2020 and $46\%$ in March 2023. 

% \begin{table}[ht]
%     \centering
%     \begin{tabular}{l|cc}
%     \hline
% Variable & Observed  & Missing  \\
% \hline

% \hline
%     \end{tabular}
%     \caption{Population summary in New York state, 2020-03}
%     \label{tab:my_label}
% \end{table}

\subsection{Results}
\begin{figure}
    \centering
    \includegraphics[width=0.5\linewidth]{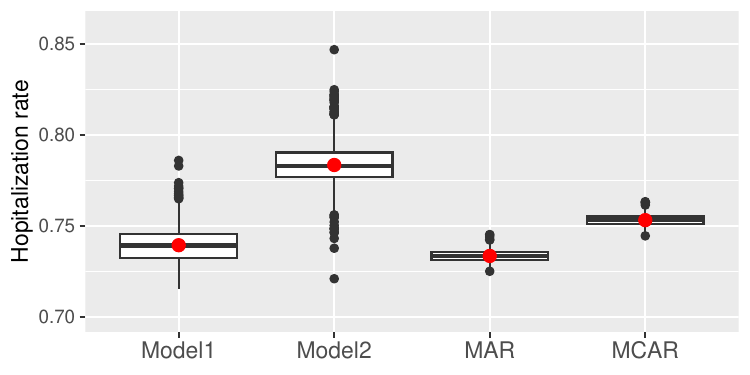}
    \vspace{-2mm}
    \caption{Boxplot of Bootstrap results (size = $1000$) using IPW estimator in Model 1, IPW estimator in Model 2, MAR estimator, and MCAR estimator. The blue dash line shows the MCAR estimate: 0.7533. The statistical summary is calculated in Table~\ref{tab:add_covid} in the Appendix.}
    \vspace{-5mm}
    \label{fig:covid}
\end{figure}

It is challenging to determine whether the cause of missing data is tied to race or hospitalization status in this COVID-19 case report. Therefore, we investigated both the IPW estimator of Model 1 and the IPW estimator of Model 2 within the dataset.  Additionally, we conducted a comparative analysis with Missing at Random (MAR) estimator and Missing Completely at Random (MCAR) estimator. Results using bootstrap with 1000 samplings are illustrated in Fig.~\ref{fig:covid}. In Model 1, When attributing missing data to race, the estimated hospitalization rate (Est: $0.7396$, 95\% CI: [$0.7190, 0.7570$]) was found to be lower than the estimate ($0.7533$) derived from the MCAR analysis. Conversely, in Model 2, assuming that missing mechanisms were linked to hospitalization resulted in an estimated rate (Est: $0.7836$, 95\% CI: [$0.7558 ,0.8071$]) surpassing the MCAR analysis. Same as Model 1, MAR also showed the MCAR result was overestimated (Est: $0.7337$, 95\% CI: [$0.7277$,$0.7395$]).
The significant differences in hospitalization rate estimates using the more naive complete case and MAR approaches compared to estimates using model 1 and model 2 underscores the importance of modeling choice in handling missing data in a principled way. 

\section{Discussion}
%\vspace{-2mm}
\label{sec:discussion}
In this paper, we introduced a data fusion approach to identification in settings where data is missing not at random (MNAR), but an auxiliary data that is missing at random (MAR) is available.

We provided identification results under two complementary models in this setting, as well as 
a straightforward-to-implement Inverse Probability Weighting (IPW) estimators for the identified parameters in each model.
We illustrated the consistency of our estimator via a simulation study.
To our knowledge, ours is the first adoption of data fusion ideas for obtaining identification in settings where data is missing not at random (MNAR).
%\zixiao{
We also applied both model 1 and model 2 to the COVID-19 case data in New York State to estimate the hospitalization rate in March 2020.
%}
A natural extension of our approach is the development of a semiparametrically efficient estimator under our proposed two models.  We leave this extension to future work, as obtaining such an estimator is nontrivial.  This is because both of our proposed models impose complex restrictions on the observed data tangent space, in addition to conditional independences implied by the graph.

\newpage
\bibliographystyle{apalike}
\bibliography{references}

% If you have textual supplementary material
\appendix

\section{PROOFS}
\label{supp:proofs}

 \begin{thma}{\ref{thm:identification1}}
 Under Assumptions~\ref{ass1:mr_g1},~\ref{ass2:mg_x}, and~\ref{ass3:yr_xmg2}, parameter $\beta = \mathbb{E} [Y^{(1)}|G =1]$ is identified using the following functional
\begin{align*}
&\mathbb{E}[Y^{(1)}|G=1]\\
&= \mathbb{E}[\mathbb{E}[\mathbb{E}[Y^{(1)}|X, G=1, M^{(1)}]|X, G=1]|G=1]\\
&= \sum_{y,m,x} y \cdot p_Y(Y^{(1)}=y|X=x, G=1, M^{(1)}=m)p_M(M^{(1)}=m|X=x, G=1)p(X=x|G=1) \\
&=  \sum_{y,m,x} y\cdot p_Y(Y^{(1)}=y|X=x, G=1, M^{(1)}=m, R=1) p_M(M^{(1)}=m|X=x, G=2)p(X=x|G=1)\\
&=  \sum_{y,m,x} y\cdot p_Y(y|X=x, G=1, M=m, R=1) p_M(m|X=x, G=2, R=1)p(X=x|G=1) \\
&= \mathbb{E}[\mathbb{E}[g_1(x,m)  \mid X=x,G=2, R=1 ]\mid G=1]\\
\end{align*}
\end{thma}

Here, equality 1 holds by the Law of Iterated Expectation,
equality 2 is by definition, equality 3 employed both Assumption~\ref{ass2:mg_x} and Assumption~\ref{ass3:yr_xmg2}, and equality 4 follows Assumption~\ref{ass1:mr_g1}.

\begin{thma}{\ref{thm:identification2}}
Under Assumption~\ref{ass1:mr_g1},~\ref{ass2:mg_x},and ~\ref{ass4}, using equation (\ref{equ:y1_r0_x1}), we have
\begin{align}
& \mathbb{E}[Y^{(1)}|G=1] \notag\\
&= \mathbb{E}[\mathbb{E}[\mathbb{E}[Y^{(1)}|M^{(1)},X,R, G=1]|X, R,G=1]|G=1] \notag\\
&= \sum_{y,m,x,r} y p(Y^{(1)}=y|X=x,R=r,G=1, M^{(1)}=m) p(M^{(1)}=m|X=x, R=r, G=1)p(X=x,R=r|G=1)  \notag\\
&= \sum_{y,m,x} y p(Y^{(1)}=y|M^{(1)}=m,X=x,R=1,G=1) p(M^{(1)}=m|X=x, R=1, G=1)p(X=x,R=1|G=1)  \notag\\
&+\sum_{y,m,x} y p(Y^{(1)}=y|M^{(1)}=m,X=x,R=0,G=1) p(M^{(1)}=m|X=x, R=0, G=1)p(X=x,R=0|G=1)  \notag\\
&= \sum_{y,m,x} y p(Y^{(1)}=y|M^{(1)}=m,X=x,R=1,G=1) p(M^{(1)}=m|X=x, R=1, G=1)p(X=x,R=1|G=1)  \notag\\
&+\sum_{y,m,x} y \frac{p\left(Y^{(1)}=y \mid M^{(1)}=m, X=x,  R=1,G=1\right) OR\left(X=x, Y^{(1)}=y\right)}{\mathbb{E}\left[OR\left(X, Y^{(1)}\right) \mid R=1,X=x,M^{(1)}=m,G=1\right]} \notag\\
&\times \frac{p(M^{(1)}=m|X=x,R=1, G=2) -p(M^{(1)}=m,R=1|X=x, G=1)}{p(R=0|X=x, G=1)}\notag \\
&\times p(X=x,R=0|G=1)  \notag \\
&= \sum_{y,m,x} y p(Y^{(1)}=y|X=x,R=1,G=1, M^{(1)}=m) \notag \\
&\times \Biggl(
p(M^{(1)}=m,X=x, R=1 | G=1)  \notag\\
&+  \frac{ OR\left(X=x, Y^{(1)}=y\right)}{\mathbb{E}\left[OR\left(X, Y^{(1)}\right) \mid R=1,X=x,M^{(1)}=m,G=1\right]} \notag\\
&\times \Bigr[\frac{p(M^{(1)}=m|X=x,R=1, G=2)}{p(R=0|X=x, G=1)} -\frac{p(M^{(1)}=m,R=1|X=x,  G=1)}{p(R=0|X=x, G=1)}\Bigr]\times p(X=x,R=0|G=1)  \notag
\Biggl)
\end{align}
\end{thma}

where the solution of $\widetilde{OR}(X,Y^{(1)})$ in equation (\ref{equ:ortilde}) is guaranteed by
Assumption~\ref{ass5:completeness}, therefore the solution for $OR(X,Y^{(1)})$ exists by definition.

Here, equality 1 follows by the Law of Iterated Expectation, equality 2 is the definition, equality 3 expands for separating cases $R={0,1}$, equality 4 uses equation (\ref{equ:y1_r0_x1}) and equation (\ref{equ:fm_x_r0_g2}) to denote the missing distribution accordingly.

Note that equation (\ref{equ:ortilde}) can be written as
\begin{align}
&\mathbb{E}\{\widetilde{OR}\left(X, Y^{(1)}\right)  \mid R=1,X,M^{(1)},G=1\} \notag\\
&=\frac{p\left(M^{(1)} \mid X, R=0, G=1\right)}{p\left(M^{(1)} \mid X, R=1, G=1\right)}\notag \\
&=\frac{p(M|X,R=1, G=2) -p(M|X, R=1, G=1)p(R=1|X, G=1)}{p(R=0|X, G=1)p\left(M^{(1)} \mid X, R=1, G=1\right)}\notag 
\end{align}
by equation (\ref{equ:fm_x_r0_g2}), which stands since
\begin{align}
&p(M^{(1)}|X, R=0, G=1)p(R=0|X, G=1)\notag\\
&=p(M^{(1)}|X, G=1)-p(M^{(1)}|X, R=1, G=1)p(R=1|X, G=1)\notag\\
&=p(M^{(1)}|X, G=2)-p(M^{(1)}|X, R=1, G=1)p(R=1|X, G=1)\notag\\
&=p(M^{(1)}|X,R=1, G=2)-p(M^{(1)}, R=1|X, G=1) \notag\\
\end{align}
Here, equality 1 is by definition as $R$ is a binary indicator, equality 2 is by Assumption \ref{ass2:mg_x}, and equality 3 is by Assumption \ref{ass1:mr_g1}.

\begin{propa}{\ref{prop:model2}}
[\cite{miao2015identification}]
In the interest of being self-contained, we proved here under Assumption~\ref{ass4}, for all $(X, Y^{(1)}, M^{(1)})$ in Domain 1, we have the following properties, 

For equation (\ref{equ:or}), we note that
\begin{align}
    OR\left(X, Y^{(1)}, M^{(1)}\right)&=\frac{p\left(Y^{(1)} \mid R=0, X, M^{(1)},G=1\right) p\left(Y^{(1)}=0 \mid R=1, X, M^{(1)},G=1\right)}{p\left(Y^{(1)} \mid R=1, X, M^{(1)},G=1\right) p\left(Y^{(1)}=0 \mid R=0, X, M^{(1)},G=1\right)}\notag\\
    &= \frac{p\left(Y^{(1)} , R=0, X, M^{(1)},G=1\right) p\left(Y^{(1)}=0, R=1, X, M^{(1)},G=1\right)}{p\left(Y^{(1)}, R=1, X, M^{(1)},G=1\right) p\left(Y^{(1)}=0, R=0, X, M^{(1)},G=1\right)}\notag\\
    &= \frac{p\left(R=0 \mid Y^{(1)} , X, M^{(1)},G=1\right) p\left(R=1\mid Y^{(1)}=0, X, M^{(1)},G=1\right)}{p\left(R=1 \mid Y^{(1)}, X, M^{(1)},G=1\right) p\left(R=0 \mid Y^{(1)}=0, X, M^{(1)},G=1\right)}\notag\\
    &= \frac{p\left(R=0 \mid Y^{(1)} , X,G=1\right) p\left(R=1\mid Y^{(1)}=0, X,G=1\right)}{p\left(R=1 \mid Y^{(1)}, X,G=1\right) p\left(R=0 \mid Y^{(1)}=0, X,G=1\right)} \notag \\
    &=   OR\left(X, Y^{(1)}\right) \notag
\end{align}

Equation (\ref{equ:y1_r0_x1}) follows by observing that
\begin{align}
    &\mathbb{E}\left[OR\left(X, Y^{(1)}\right) \mid R=1,X,M^{(1)},G=1
    \right]\notag\\
    &=\mathbb{E}\left[
     \frac{p\left(R=0 \mid Y^{(1)} , X, M^{(1)},G=1\right) p\left(R=1\mid Y^{(1)}=0, X, M^{(1)},G=1\right)}{p\left(R=1 \mid Y^{(1)}, X, M^{(1)},G=1\right) p\left(R=0 \mid Y^{(1)}=0, X, M^{(1)},G=1\right)}
    \mid R=1,X,M^{(1)},G=1
    \right]\notag\\
    & = \frac{p\left(R=1 \mid Y^{(1)}=0, X,M^{(1)},G=1\right)}{p\left(R=0 \mid Y^{(1)}=0, X,M^{(1)},G=1\right)}\mathbb{E}\left[\frac{p\left(R=0 \mid Y^{(1)}, X,M^{(1)},G=1\right)}{p\left(R=1 \mid Y^{(1)},X,M^{(1)}, G=1\right)}\mid R=1,X,M^{(1)},G=1\right]\notag\\
    & = \frac{p\left(R=1 \mid Y^{(1)}=0, X,M^{(1)},G=1\right)p\left(R=0, X,M^{(1)},G=1\right)}{p\left(R=0 \mid Y^{(1)}=0, X,M^{(1)},G=1\right)p\left(R=1, X,M^{(1)},G=1\right)} \notag\\
    &  \underbrace{\mathbb{E}\left[\frac{p\left( Y^{(1)} \mid R=0 , X,M^{(1)},G=1\right)}{p\left( Y^{(1)} \mid R=1 , X,M^{(1)},G=1\right)} \mid R=1,X,M^{(1)},G=1\right]}_{=1}\notag\\
    & = \frac{p\left(Y^{(1)}=0\mid  R=1,X,M^{(1)},G=1\right)}{p\left(Y^{(1)}=0 \mid  R=0,X,M^{(1)},G=1\right)} \notag
\end{align}

For equation (\ref{equ:model2_r1_xy1g2}),
\begin{align}
&p\left(R=1 \mid X, Y^{(1)}, G=1\right)^{-1} \notag \\
&=\frac{1}{p\left(R=1 \mid Y^{(1)}, X,G=1\right)}\notag\\
&=  \frac{p\left(R=0 \mid Y^{(1)} , X,G=1\right)+p\left(R=1 \mid Y^{(1)}, X,G=1\right) }{p\left(R=1 \mid Y^{(1)}, X,G=1\right) }     \notag\\
&=   1+ \frac{p\left(R=0 \mid Y^{(1)} , X,G=1\right) }{p\left(R=1 \mid Y^{(1)}, X,G=1\right) }   \notag\\
&=1 + \frac{p\left(R=0 \mid Y^{(1)} , X,G=1\right) p\left(R=1\mid Y^{(1)}=0, X,G=1\right)}{p\left(R=1 \mid Y^{(1)}, X,G=1\right) p\left(R=0 \mid Y^{(1)}=0, X,G=1\right)} \frac{ p\left(R=0 \mid X, Y^{(1)}=0, G=1\right)}{p\left(R=1 \mid X, Y^{(1)}=0, G=1\right)}\notag\\
&= 1 + \frac{OR\left(X, Y^{(1)}\right) p\left(R=0 \mid X, Y^{(1)}=0, G=1\right)}{p\left(R=1 \mid X, Y^{(1)}=0, G=1\right)}\notag
\end{align}
Equation (\ref{equ:r1_x_y0_g2}) stands as we first observed 
\begin{equation}
\begin{aligned}
    &\mathbb{E}\left[OR\left(X, Y^{(1)}\right) \mid R=1,X,G=1
    \right]\\
    &=\mathbb{E}\left[
    \frac{p\left(R=0 \mid Y^{(1)} , X,G=1\right) p\left(R=1\mid Y^{(1)}=0, X,G=1\right)}{p\left(R=1 \mid Y^{(1)}, X,G=1\right) p\left(R=0 \mid Y^{(1)}=0, X,G=1\right)}
    \mid R=1,X,G=1
    \right]\\
    & = \frac{p\left(R=1 \mid Y^{(1)}=0, X,G=1\right)}{p\left(R=0 \mid Y^{(1)}=0, X,G=1\right)}\mathbb{E}\left[\frac{p\left(R=0 \mid Y^{(1)} , X,G=1\right)}{p\left(R=1 \mid Y^{(1)}, X,G=1\right)}\mid R=1,X,G=1\right]\\
    & = \frac{p\left(R=1 \mid Y^{(1)}=0, X,G=1\right)p\left(R=0, X,G=1\right)}{p\left(R=0 \mid Y^{(1)}=0, X,G=1\right)p\left(R=1, X,G=1\right)} \notag\\
    & \underbrace{\mathbb{E}\left[\frac{p\left( Y^{(1)} \mid R=0 , X,G=1\right)}{p\left( Y^{(1)} \mid R=1 , X,G=1\right)} \mid R=1,X,G=1\right]}_{=1}\\
    & = \frac{p\left(Y^{(1)}=0\mid  R=1,X,G=1\right)}{p\left(Y^{(1)}=0 \mid  R=0,X,G=1\right)}
\end{aligned}
\end{equation}

Then we note that
\begin{align}
&\frac{\mathbb{E}\left[OR\left(X, Y^{(1)}\right) \mid R=1,X,G=1\right]}{\mathbb{E}\left[OR\left(X, Y^{(1)}\right) \mid R=1,X,G=1\right] + p(R=0 \mid X, G=1)/p(R=1 \mid X,G=1) }\notag\\
&=\frac{\frac{p\left(Y^{(1)}=0\mid  R=1,X,G=1\right)}{p\left(Y^{(1)}=0 \mid  R=0,X,G=1\right)}}{\frac{p\left(Y^{(1)}=0\mid  R=1,X,G=1\right)}{p\left(Y^{(1)}=0 \mid  R=0,X,G=1\right)}+\frac{p(R=0,X, G=1)}{p(R=1,X,G=1)} }\notag\\
&= \frac{p\left(Y^{(1)}=0,R=1,X,G=1\right)}{p\left(Y^{(1)}=0,  R=1,X,G=1\right)+p\left(Y^{(1)}=0, R=0,X,G=1\right)}\notag\\
& = \frac{ p\left(R=1 \mid Y^{(1)}=0,X,G=1\right)}{p\left(R=1 \mid Y^{(1)}=0,X,G=1\right)+p\left(R=0 \mid Y^{(1)}=0, X,G=1\right)}\notag\\
&= p\left(R=1 \mid Y^{(1)}=0,X,G=1\right)\notag
\end{align}

Equation (\ref{equ:ortilde}) is true because
\begin{align}
&\widetilde{OR}\left(X, Y^{(1)}\right)\notag\\
&= \frac{OR\left(X, Y^{(1)}\right)}{{\mathbb{E}\{OR\left(X, Y^{(1)}\right) \mid R=1,X,G=1\}}} \notag\\
&= \frac{\frac{p\left(R=0 \mid Y^{(1)} , X,G=1\right) p\left(R=1\mid Y^{(1)}=0, X,G=1\right)}{p\left(R=1 \mid Y^{(1)}, X,G=1\right) p\left(R=0 \mid Y^{(1)}=0, X,G=1\right)}}{\frac{p\left(Y^{(1)}=0\mid  R=1,X,G=1\right)}{p\left(Y^{(1)}=0 \mid  R=0,X,G=1\right)}}\notag\\
& = \frac{p\left(Y^{(1)}\mid R=0,X,G=1\right)}{p\left(Y^{(1)} \mid R=1,X,G=1\right)} \notag
\end{align}
So by leverageing Assumption~\ref{ass4}, we have
\begin{align}
&\mathbb{E}\left[\widetilde{OR}\left(X, Y^{(1)}\right)  \mid R=1,X,M^{(1)},G=1\right]= \mathbb{E}\left[
\frac{p\left(Y^{(1)}\mid R=0,X,G=1\right)}{p\left(Y^{(1)} \mid R=1,X,G=1\right)} \mid R=1,X,M^{(1)},G=1
\right]\notag \\
&= \frac{p\left(M^{(1)} \mid X, R=0 ,G=1\right) }{p\left(M^{(1)} \mid X, R=1 ,G=1\right) } \underbrace{\mathbb{E}\left[
\frac{p\left(Y^{(1)}\mid M^{(1)},R=0,X,G=1\right)}{p\left(Y^{(1)} \mid M^{(1)},R=1,X,G=1\right)} \mid R=1,X,M^{(1)},G=1
\right]}_{=1}\notag 
\end{align}

\end{propa}

\begin{propa}{\ref{prop:model1.ipw}}

we first prove the lemma below:

\begin{lem}
Define $q(X,Y^{(1)})=1/p(R=1|X,M^{(1)},G=1)$, for any specific functin $g(X,M^{(1)},Y^{(1)})$, under Assumption~\ref{ass3:yr_xmg2},
\begin{align}
    \mathbb{E}\left[ \left( q(X,M^{(1)})R-1  \right) g(X,M^{(1)},Y^{(1)})\mid G=1\right]=0
    \label{equ:lemmaq}
\end{align}
\label{lem:q}
\end{lem}

To prove equation (\ref{equ:lemmaq}), we first notice that 

\begin{align}
&\mathbb{E}\left[
\left(q(X,M^{(1)})R-1  \right) g(X,M^{(1)},Y^{(1)})
\mid X, M^{(1)},G=1
\right] \notag \\
& =\underbrace{\mathbb{E}\left[
\left(q(X,M^{(1)})R-1  \right) 
\mid X, M^{(1)},G=1
\right]}_{=0}\cdot 
\mathbb{E}\left[
g(X,M^{(1)},Y^{(1)} 
\mid X, M^{(1)},G=1
\right]\notag\\
&=0\notag
\end{align}
Using the Law of Iterated Expectation, we have
\begin{align}
&\mathbb{E}\left[ \left( q(X,M^{(1)})R-1  \right) g(X,M^{(1)},Y^{(1)})\mid G=1\right]\notag\\
&\mathbb{E}\left[\mathbb{E}\left[ \left( q(X,M^{(1)})R-1  \right) g(X,M^{(1)},Y^{(1)})\mid X, M^{(1)},G=1\right]\mid G=1\right]\notag\\
&=0\notag
\end{align}

Noticing $\mathbb{E}[Y^{(1)}- \beta \mid G=1]=0$ by definition, let $g(X,M^{(1)},Y^{(1)})=Y^{(1)}- \beta$ in equation (\ref{equ:lemmaq}),
\begin{equation}
\begin{aligned}
    &\mathbb{E}\left[ (q(X,M^{(1)};\hat{\alpha})R -1) \cdot (Y^{(1)}- \beta) |G=1\right] \notag \\
     &= \mathbb{E}\left[ (q(X,M^{(1)};\hat{\alpha})R\cdot Y^{(1)}   - \beta|G=1\right]  \notag \\
     &=0 \notag
\end{aligned}
\end{equation}

Additionally, let $g(X,M^{(1)},Y^{(1)})=h(X,M^{(1)})$, where $h(X,M^{(1)})$ is any specific function of $\left( X,M^{(1)}\right)$, under Assumption \ref{ass1:mr_g1} and \ref{ass2:mg_x}, we have

\begin{align}
&\mathbb{E}\left[ \left( w(X,M^{(1)})R-1  \right) h(X,M^{(1)})|G=1\right] \notag\\
& =\mathbb{E}\left[ \left( w(X,M^{(1)})R \cdot h(X,M^{(1)})  \right)|G=1\right] - \mathbb{E}\left[  h(X,M^{(1)})|G=1\right] \notag\\
& = \mathbb{E}\left[ \left( w(X,M^{(1)})R \cdot h(X,M^{(1)})  \right)|G=1\right] - \sum_{X,M^{(1)}} h(X,M^{(1)}) p(M^{(1)} \mid X,G=1) p(X \mid G=1) \notag\\
& = \mathbb{E}\left[ \left( w(X,M^{(1)})R \cdot h(X,M^{(1)})  \right)|G=1\right] - \sum_{X,M^{(1)}} h(X,M^{(1)}) p(M^{(1)} \mid X,G=2) p(X \mid G=1) \notag\\
&= \mathbb{E}\left[ \left( w(X,M^{(1)})R \cdot h(X,M^{(1)})  \right)|G=1\right] - \sum_{X,M^{(1)}} h(X,M^{(1)}) p(M^{(1)} \mid X, R=1,G=2) p(X \mid G=1) \notag\\
&= \mathbb{E}\left[ \left( w(X,M^{(1)})R \cdot h(X,M^{(1)})  \right) -  \mathbb{E}\left[ h(X,M) \mid X, R=1, G=2 \right] \mid G=1\right] \notag\\
&=0 \notag
\end{align}

\end{propa}

Proof of IPW estimation approach for Model 2 relies on the following lemma, as stated in \citep{miao2015identification}:

\begin{propa}{\ref{prop:model2.estimation}}we first prove the lemma below:
\begin{lem}
Define $w(X,Y^{(1)})=1/p(R=1|X,Y^{(1)},G=1)$, for any specific functin $g(X,M^{(1)},Y^{(1)})$,  under Assumption~\ref{ass4}, we have
\begin{align}
    \mathbb{E}\left[ \left( w(X,Y^{(1)})R-1  \right) g(X,M^{(1)},Y^{(1)})|G=1\right]=0
    \label{equ:lemmaw}
\end{align}
\label{lem:w}
\end{lem}

To prove equation (\ref{equ:lemmaw}), we first notice

\begin{align}
&\mathbb{E}\left[
\left(w(X,Y^{(1)})R-1  \right) g(X,M^{(1)},Y^{(1)})
\mid X, Y^{(1)},G=1
\right] \notag \\
& =\underbrace{\mathbb{E}\left[
\left(w(X,Y^{(1)})R-1  \right) 
\mid X, Y^{(1)},G=1
\right]}_{=0}\cdot 
\mathbb{E}\left[
g(X,M^{(1)},Y^{(1)} 
\mid X, Y^{(1)},G=1
\right]\notag\\
&=0\notag
\end{align}
Using the Law of Iterated Expectation, we have
\begin{align}
&\mathbb{E}\left[ \left( w(X,Y^{(1)})R-1  \right) g(X,M^{(1)},Y^{(1)})\mid G=1\right]\notag\\
&\mathbb{E}\left[\mathbb{E}\left[ \left( w(X,Y^{(1)})R-1  \right) g(X,M^{(1)},Y^{(1)})\mid X, Y^{(1)},G=1\right]\mid G=1\right]\notag\\
&=0\notag
\end{align}

Noticing $\mathbb{E}[Y^{(1)}-\beta \mid G=1]=\mathbb{E}[Y^{(1)}- \beta \mid G=1]=0$, let $g(X,M^{(1)},Y^{(1)})=Y^{(1)}- \beta$ in equation (\ref{equ:lemmaw}),
\begin{equation}
\begin{aligned}
    0 &=\mathbb{E}\left[ (w(X,Y^{(1)};\hat{\alpha},\hat{\gamma})R -1) \cdot U(\beta) |G=1\right] \\
     &= \mathbb{E}\left[ (w(X,Y^{(1)};\hat{\alpha},\hat{\gamma})R\cdot Y^{(1)}   - \beta|G=1\right] 
\end{aligned}
\end{equation}
Also, let $g(X,M^{(1)},Y^{(1)})=h(X,M^{(1)})$, where $h(X,M^{(1)})$ is any specific function of $(X,M^{(1)})$, under Assumption \ref{ass1:mr_g1} and \ref{ass2:mg_x}, we have
\begin{align}
0 & =\mathbb{E}\left[ \left( w(X,Y^{(1)})R-1  \right) h(X,M^{(1)})|G=1\right]\notag\\
& =\mathbb{E}\left[ \left( w(X,Y^{(1)})R \cdot h(X,M^{(1)})  \right)|G=1\right] - \mathbb{E}\left[  h(X,M^{(1)})|G=1\right]\notag\\
& = \mathbb{E}\left[ \left( w(X,Y^{(1)})R \cdot h(X,M^{(1)})  \right)|G=1\right] - \sum_{X,M^{(1)}} h(X,M^{(1)}) p(M^{(1)} \mid X,G=1) p(X \mid G=1)\notag\\
& = \mathbb{E}\left[ \left( w(X,Y^{(1)})R \cdot h(X,M^{(1)})  \right)|G=1\right] - \sum_{X,M^{(1)}} h(X,M^{(1)}) p(M^{(1)} \mid X,G=2) p(X \mid G=1)\notag\\
&= \mathbb{E}\left[ \left( w(X,Y^{(1)})R \cdot h(X,M^{(1)})  \right)|G=1\right] - \sum_{X,M^{(1)}} h(X,M^{(1)}) p(M^{(1)} \mid X, R=1,G=2) p(X \mid G=1)\notag\\
&= \mathbb{E}\left[ \left( w(X,Y^{(1)})R \cdot h(X,M^{(1)})  \right) -  \mathbb{E}\left[ h(X,M) \mid X, R=1, G=2 \right] \mid G=1\right]\notag
\end{align}
\end{propa}

\section{DATA GENERATION AND ESTIMATION}

\subsection{Model 1}
We create a binary grouping variable, denoted as $G$ and taking values from the set $\{1, 2\}$, for a dataset with a total sample size of $n$, assuming that each $G$ follows a Bernoulli distribution with parameters $0.5$.

For the primary domain, we assume the data generated as follows satisfying Assumption~\ref{ass3:yr_xmg2}:
\begin{align}
&X \mid G=1 \sim \mathcal{N}\left(1,1\right) \notag\\
&M^{(1)} \mid X, G=1  \sim \mathcal{N}\left(\beta_{m 0}+\beta_{m 1} X+\beta_{m 2} X^{2}, 1\right) \notag \\
&Y^{(1)} \mid X, M^{(1)}, G=1  \sim \mathcal{N}\left(\beta_{y 0}+\beta_{y 1} X+\beta_{y 2} X^{2}+\beta_{y 3} M^{(1)}, 1\right) \notag \\
&p(R=1\mid X,M^{(1)},G=1)= logis(a_0 + a_1X + a_2M^{(1)}+a_3[M^{(1)}]^2) \notag
\end{align}

For the correctly specified model (T), we assume $a_3=0$, and for the misspecified setting, we have $a_3=-1$ in particular.

The auxiliary domain is generated so as to satisfy Assumption~\ref{ass1:mr_g1} and \ref{ass2:mg_x}:
\begin{align}
&X \mid G=2 \sim \mathcal{N}\left(0,1\right) \notag\\
&M^{(1)} \mid X, G=2  \sim \mathcal{N}\left(\beta_{m 0}+\beta_{m 1} X+\beta_{m 2} X^{2}, 1\right) \notag \\
&p(R=1\mid X,M^{(1)},G=2)= logis(b_0 + b_1X) \notag
\end{align}
The parameters used for presenting results have been discussed in the paper. Details of the simulation setting can be found in the code scripts.

\subsection{Model 2}
For a dataset with total sample size $n$, we first generate the binary grouping variable $G$, $G\in \{1,2   \}$,by assuming each $G \sim Bernouli(0.5)$

In the primary domain ($G=1$), the data is generated by assuming:

\begin{align}
&X \mid G=1 \sim \mathcal{N}(0,1) \notag\\
&M^{(1)} \mid R=1,X,G=1  \sim \mathcal{N}\left(\beta_{m 0}+\beta_{m 1} X+\beta_{m 2} X^{2}, 1\right)\notag\\
&Y^{(1)} \mid R=1, X, M^{(1)},G=1 \sim \mathcal{N}\left(\beta_{y 0}+\beta_{y 1} X+\beta_{y 2} X^{2}+\beta_{y 3} M^{(1)}, 1\right)\notag\\
&\operatorname{logit} p(R=1 \mid X, Y^{(1)}=0,G=1)  =\alpha_{0}+\alpha_{1} X+\alpha_{2} X^{2}\notag\\
&\operatorname{OR}(X, Y^{(1)}) =\exp (-\gamma Y^{(1)}) \notag
\end{align}

Overall speaking, we generate the dataset using following the variable order: $X \rightarrow R\rightarrow M \rightarrow Y$. To achieve such a data generation order, we leverage the proposed properties in Proposition~\ref{prop:model2} under Assumption~\ref{ass3:yr_xmg2}.

We first calculated some functions that will be used in the generating process:

For $\mathbb{E}[\operatorname{OR}(X, Y^{(1)}) \mid R=1, X, M] $, using the above proposed models, let $\mu_{Y}  =\beta_{y 0}+\beta_{y 1} X+\beta_{y 2} X^{2}+\beta_{y 3} M^{(1)} $, then we have
\begin{equation}
\begin{aligned}
\mathbb{E}[\operatorname{OR}(X, Y^{(1)}) \mid R=1, X, M,G=1 ] & =\frac{1}{\sqrt{2\pi}} \int \exp \left(-\gamma y-\frac{1}{2}\left(y-\mu_{Y}\right)^{2}\right) dy \\
& =\frac{1}{\sqrt{2\pi}} \int \exp \left(-\frac{1}{2}\left(y-\mu_{Y}+\gamma\right)^{2}-\frac{1}{2}\left(2 \mu_{Y} \gamma-\gamma^{2}\right)\right) dy \\
& =\exp \left(-\mu_{Y} \gamma+\frac{1}{2} \gamma^{2}\right) \notag
\end{aligned}
\end{equation}

In a similar way, with regard to the conditional expectation $\mathbb{E}[\operatorname{OR}(X, Y) \mid R=1, X,G=1 ]$, let  $\tilde{\mu}{Y} = \beta_{y0} + \beta_{y1}X + \beta_{y2}X^2$ and $\mu_{M} = \beta_{z0} + \beta_{z1}X + \beta_{z2}X^2$, we can establish the following functional relationships utilizing the previously mentioned equation:

\begin{equation}
\begin{aligned}
\mathbb{E}[\operatorname{OR}(X, Y^{(1)}) \mid R=1, X,G=1 ] & =\mathbb{E}[\mathbb{E}[\operatorname{OR}(X, Y^{(1)})  \mid R=1, X,M,G=1 ]\mid R=1, X,G=1 ]\\
& = \frac{1}{\sqrt{2\pi}} \int \exp \left(-\mu_{Y} \gamma+\frac{1}{2} \gamma^{2}-\frac{1}{2}\left(m-\mu_{M}\right)^{2}\right) dm \\
& =\exp \left(-\tilde{\mu}_{Y} \gamma+\frac{1}{2} \gamma^{2}\right) \times \frac{1}{\sqrt{2}} \int \exp \left(-\gamma \beta_{y 3} m-\frac{1}{2}\left(m-\mu_{M}\right)^{2}\right) d m \\
& =\exp \left(-\tilde{\mu}_{Y} \gamma+\frac{1}{2} \gamma^{2}-\mu_{M} \beta_{y 3} \gamma+\frac{1}{2} \beta_{y 3}^{2} \gamma^{2}\right)\notag
\end{aligned}
\end{equation}

Having assembled the essential components as outlined in the equation (\ref{equ:r1_x_y0_g2})
we can obtain the distribution of 
$R=1 \mid X, G=1$.
Let $\mu_{R}=\alpha_{0}+\alpha_{1} X+\alpha_{2} X^{2}$, we rewrite equation (\ref{equ:r1_x_y0_g2}) to be

\begin{align}
    p(R=1 \mid X, Y^{(1)}=0,G=1) & =\frac{\mathbb{E}[\operatorname{OR}(X, Y^{(1)}) \mid R=1, X,G=1]}{p(R=0 \mid X,G=1) / p(R=1 \mid X,G=1)+\mathbb{E}[\operatorname{OR}(X, Y^{(1)}) \mid R=1, X,G=1]} \notag
\end{align}

Therefore,
\begin{align}
\frac{p(R=0 \mid X,G=1)}{p(R=1 \mid X,G=1)} & =\mathbb{E}[\operatorname{OR}(X, Y^{(1)}) \mid R=1, X,G=1] \times \frac{1-p(R=1 \mid X, Y^{(1)}=0,G=1)}{p(R=1 \mid X, Y^{(1)}=0,G=1)} \notag\\
& =\exp \left(-\tilde{\mu}_{Y} \gamma+\frac{1}{2} \gamma^{2}-\mu_{M} \beta_{y 3} \gamma+\frac{1}{2} \beta_{y 3}^{2} \gamma^{2}\right) \times \exp \left(-\alpha_{0}-\alpha_{1} X-\alpha_{2} X^{2}\right) \notag\\
& =\exp \left(-\tilde{\mu}_{Y} \gamma+\frac{1}{2} \gamma^{2}-\mu_{M} \beta_{y 3} \gamma+\frac{1}{2} \beta_{y 3}^{2} \gamma^{2}-\mu_{R}\right)\notag
\end{align} 

so that $\operatorname{logit} p(R=1 \mid X,G=1)=\tilde{\mu}_{Y} \gamma-\frac{1}{2} \gamma^{2}+\mu_{M} \beta_{y 3} \gamma-\frac{1}{2} \beta_{y 3}^{2} \gamma^{2}+\mu_{R}$

Then we calculate for $M^{(1)} | R = 0, X,G=1$ using the combination of equations (\ref{equ:ortilde}) and (\ref{equ:ortildeor}), 

\begin{align}
&p(M^{(1)} \mid R=0, X,G=1) \notag\\ 
&=p(M^{(1)} \mid R=1, X,G=1) \times \frac{\mathbb{E}(\mathrm{OR}(X, Y^{(1)}) \mid R=1, X, M^{(1)},G=1)}{\mathbb{E}(\mathrm{OR}(X, Y^{(1)}) \mid R=1, X,G=1)} \notag\\
& =\frac{1}{\sqrt{2\pi}} \exp \left(-\frac{1}{2}\left(M^{(1)}-\mu_{M}\right)^{2}\right) \exp \left(-\mu_{Y} \gamma+\frac{1}{2} \gamma^{2}-\left(-\tilde{\mu}_{Y} \gamma+\frac{1}{2} \gamma^{2}-\mu_{M} \beta_{y 3} \gamma+\frac{1}{2} \beta_{y 3}^{2} \gamma^{2}\right)\right) \notag\\
& =\frac{1}{\sqrt{2\pi}} \exp \left(-\frac{1}{2}\left(M^{(1)}-\mu_{M}\right)^{2}\right) \times \exp \left(-\beta_{y 3} \gamma Z+\mu_{M} \beta_{y 3} \gamma-\frac{1}{2} \beta_{y 3}^{2} \gamma^{2}\right) \notag\\
& =\frac{1}{\sqrt{2\pi}} \exp \left(-\frac{1}{2}\left(M^{(1)}-\mu_{M}\right)^{2}-\beta_{y 3} \gamma\left(M^{(1)}-\mu_{M}\right)-\frac{1}{2} \beta_{y 3}^{2} \gamma^{2}\right)\notag \\
& =\frac{1}{\sqrt{2\pi}} \exp \left(-\frac{1}{2}\left(M^{(1)}-\mu_{M}+\beta_{y 3} \gamma\right)^{2}\right)\notag
\end{align}

So  $M^{(1)} \mid R=0, X,G=1$  is a normal distribution with mean  $\mu_{M}-\beta_{y 3} \gamma$  and variance 1 .

Upon acquiring the probability distributions for both the random variables, $R$ and $M^{(1)}$, we employ equation (\ref{equ:y1_r0_x1}) to derive the conditional probability distribution of $Y^{(1)}$ under the given conditions: $R=0, X, M^{(1)}$, and $G=1$ as follows:

\begin{equation}
\begin{aligned}
p(Y^{(1)} \mid R=0, X, M^{(1)},G=1) & 
\frac{p\left(Y^{(1)} \mid R=1, X, M^{(1)}, G=1\right) OR\left(X, Y^{(1)}\right)}{\mathbb{E}\left[OR\left(X, Y^{(1)}\right) \mid R=1,X,M^{(1)},G=1\right]}\\
& =\frac{1}{\sqrt{2\pi}} \exp \left(-\gamma y-\frac{1}{2}\left(y-\mu_{Y}\right)^{2}+\mu_{Y} \gamma-\frac{1}{2} \gamma^{2}\right) \\
& = \frac{1}{\sqrt{2\pi}} \exp  \left(-\frac{1}{2}\left(\left(y-\mu_{Y}\right)^{2}+2 \gamma(y-\mu_{Y}) +\gamma^2\right)\right)  \\
& =\frac{1}{\sqrt{2\pi}} \exp \left(-\frac{1}{2}\left( y-\mu_{Y}+\gamma \right)^2\right) \notag
\end{aligned}
\end{equation}

As a result, $Y^{(1)} \mid R=0, X, M{(1)},G=1 \sim \mathcal{N}\left(\mu_{Y}-\gamma, 1\right)$

In conclusion, we are prepared to articulate the data generation process as follows:

Denoting
\begin{align}
    &\mu_{Y}  =\beta_{y 0}+\beta_{y 1} X+\beta_{y 2} X^{2}+\beta_{y 3} M^{(1)} \notag\\
&\tilde{\mu}_{Y}  =\beta_{y 0}+\beta_{y 1} X+\beta_{y 2} X^{2} \notag\\
&\mu_{M}  =\beta_{m 0} + \beta_{m 1} X+\beta_{m 2} X^{2} \notag\\
&\mu_{R}=\alpha_{0}+\alpha_{1} X+\alpha_{2} X^{2}\notag
\end{align}

The dataset of Domain 1 is generated in a sequential order as follows:
\begin{align}
    & X \mid G=1 \sim \mathcal{N}(0,1)\notag\\ 
&\operatorname{logit} p(R=1 \mid X,G=1)  =\tilde{\mu}_{Y} \gamma-\frac{1}{2} \gamma^{2}+\mu_{M} \beta_{y 3} \gamma-\frac{1}{2} \beta_{y 3}^{2} \gamma^{2}+\mu_{R}\notag\\
& M \mid R=1, X,G=1 \sim \mathcal{N}\left(\underbrace{\beta_{m 0}+\beta_{m 1} X+\beta_{m 2} X^{2}}_{\mu_M}, 1\right)\notag\\ 
& M \mid R=0, X,G=1  \sim \mathcal{N}\left(\mu_{M}-\beta_{y 3}\gamma, 1\right)\notag\\
&Y \mid R=1, X, M^{(1)},G=1 \sim \mathcal{N}\left(\underbrace{\beta_{y 0}+\beta_{y 1} X+\beta_{y 2} X^{2}+\beta_{y 3} M}_{\mu_Y}, 1\right)\notag\\
&Y \mid R=0, X, M^{(1)},G=1 \sim \mathcal{N}\left(\mu_{Y}-\gamma, 1\right) \notag
\end{align}

After the data generation of the primary domain, we generate the data of the auxiliary domain. The key constrain of this auxiliary domain in simulation is Assumption~\ref{ass2:mg_x}. To satisfy that, we first generate a temporary variable $R_{tmp}$ in order to obtain the distribution for $M^{(1)}|X,G=2$ as follows:
\begin{align}
& X \mid G=2 \sim \mathcal{N}(1,1)\notag\\
&\operatorname{logit} p(R_{tmp}=1 \mid X,G=2)  =\tilde{\mu}_{Y} \gamma-\frac{1}{2} \gamma^{2}+\mu_{M} \beta_{y 3} \gamma-\frac{1}{2} \beta_{y 3}^{2} \gamma^{2}+\mu_{R}\notag\\
& M \mid R_{tmp}=1, X,G=2 \sim \mathcal{N}\left(\beta_{m 0}+\beta_{m 1} X+\beta_{m 2} X^{2}, 1\right)\notag\\ 
& M \mid R_{tmp}=0, X,G=2 \sim \mathcal{N}\left(\mu_{M}-\beta_{y 3}\gamma, 1\right)\notag
\end{align}
The missing mechanism only depends on $X$ and we further generate the true $R$ in Domain 2 by assuming 
\begin{align}
p(R=1 \mid X, G=2) = logis(c_0+c_1x) \notag
\end{align}
where $logis(x)=(1 + \exp(-x))^{-1}$.

For the inverse probability weighting estimation method, we need to estimate $p (R = 1 | X, Y,G=1;\alpha,\gamma)$,  as in equation (\ref{equ:model2_r1_xy1g2}). Recall that  $\operatorname{logit} p(R=1 \mid X, Y=0,G=1)  =\alpha_{0}+\alpha_{1} X+\alpha_{2} X^{2} $ ,
 $\operatorname{OR}(X, Y^{(1)}) =\exp (-\gamma Y^{(1)}) $, therefore we have
\begin{align}
    &  \frac{1}{p (R = 1 | X, Y^{(1)},G=1)}\notag\\
    & = \frac{ p (R = 1 | X, Y^{(1)} = 0,G=1;\alpha) + OR(X, Y^{(1)};\gamma )p (R = 0 | X, Y^{(1)} = 0,G=1;\alpha)}{f (R = 1 | X, Y^{(1)} = 0,G=1;\alpha)}\notag\\
    & = 1 +  OR(X, Y^{(1)};\gamma )\frac{ p(R = 0 | X, Y^{(1)} = 0,G=1;\alpha)}{p (R = 1 | X, Y^{(1)} = 0,G=1;\alpha)}\notag\\
    & = 1+ \exp(-\gamma Y^{(1)}) \exp\left(-(\alpha_{0}+\alpha_{1} X+\alpha_{2} X^{2})\right)\notag\\
    &=1+ \exp \left(-\gamma Y^{(1)}-(\alpha_{0}+\alpha_{1} X+\alpha_{2} X^{2}) \right)\notag
\end{align}
As a result, $p (R = 1 | X, Y^{(1)},G=1)= \frac{1}{1+ \exp \left(-\gamma Y^{(1)}-(\alpha_{0}+\alpha_{1} X+\alpha_{2} X^{2}) \right)}$. The parameters used for presenting results have been discussed in the paper. Details of the simulation setting can be found in the code scripts.

\subsection{Additional Simulation Results}
In this section, we discussed simulation details on true value, and statistics of results for Model 1 displayed in Fig.~\ref{fig:model1_T} and Fig.~\ref{fig:model1_F}, and for Model 2 displayed in Fig.~\ref{fig:model2_T} and Fig.~\ref{fig:model2_F}. Also, we presented the full table of bootstrapping results in Table~\ref{tab:add1} and Table~\ref{tab:add2}.

The true value of model 1 is calculated theoretically, where $\beta=1.8$ for both model 1 (T) and model 1 (F). The true value of model 2 is calculated by taking the sample mean of $Y^{(1)}$ in the primary domain in $1000$ trials of sample size $n=20000$. True value of $\beta=-0.659$ for T and $\beta=-0.615$ for F settings.

{\scriptsize
\begin{table}[h]
\caption{Results for Model 1 displayed in Fig.~\ref{fig:model1_T} and Fig.~\ref{fig:model1_F}.\\(True value: $\beta=1.8$ for both T and F settings) } 
\label{tab:box.m1}
\begin{center}
\begin{tabular}{l|lllll}
\hline
\textbf{Setting} & \textbf{Bias} & \textbf{\%Bias} & \textbf{MSE} & \textbf{Var} \\ % Retain the old column headers
\hline
IPW (n=500,T) & 0.006 & 0.003 & 0.062 & 0.034 \\ % Replace the values
IPW (n=1000,T) & 0.005 & 0.003 & 0.007 & 0.017 \\
IPW (n=2000,T) & -0.002 & -0.001 & 0.024 & 0.009 \\
MAR (n=500,T) & 0.269 & 0.149 & 0.127 & 0.025 \\
MAR (n=1000,T) & 0.270 & 0.150 & 0.095 & 0.011 \\
MAR (n=2000,T) & 0.270 & 0.150 & 0.010 & 0.006 \\
IPW (n=500,F) & -0.040 & -0.022 & 0.026 & 0.053 \\
IPW (n=1000,F) & -0.031 & -0.017 & 0.121 & 0.027 \\
IPW (n=2000,F) & -0.017 & -0.010 & 0.064 & 0.014 \\
MAR (n=500,F) & -0.544 & -0.302 & 0.244 & 0.027 \\
MAR (n=1000,F) & -0.541 & -0.300 & 0.486 & 0.013 \\
MAR (n=2000,F) & -0.544 & -0.302 & 0.403 & 0.006 \\
\hline
\end{tabular}
\end{center}
\end{table}
}
{\scriptsize
\begin{table}[ht]
\caption{Results for Model 2 displayed in Fig.~\ref{fig:model2_T} and Fig.~\ref{fig:model2_F}.\\(True value: $\beta=-0.659$ (T) , $\beta=-0.615$ (F)} 
\label{tab:box.m2}
\begin{center}
\begin{tabular}{l|lllll}
\hline
\textbf{Setting}  &\textbf{Bias} & \textbf{\%Bias}   &\textbf{MSE} & \textbf{Var}   \\
\hline %\\
IPW (n=500,T)& 0.029  &  -0.044 & 0.010 & 0.022 \\
IPW (n=1000,T) & 0.018 &  -0.027 & 0.023 & 0.010 \\
IPW (n=2000,T)  & 0.015 &  -0.023 & 0.00007 & 0.005 \\
MAR (n=500,T) & 0.363 &  -0.551 & 0.032 & 0.020 \\
MAR (n=1000,T)& 0.363 &  -0.551 & 0.241 & 0.011 \\
MAR (n=2000,T) & 0.358 &  -0.543 & 0.112 & 0.005 \\
IPW (n=500,F)& 0.087 &  -0.142 & 0.001 & 0.018 \\
IPW (n=1000,F)& 0.086 &  -0.139 & 0.010 & 0.009 \\
IPW (n=2000,F)& 0.084 &  -0.136 & 0.028 & 0.005 \\
MAR (n=500,F) & 0.195 &  -0.317 & 0.023 & 0.021 \\
MAR (n=1000,F)& 0.193 &  -0.314 & 0.044 & 0.010 \\
MAR (n=2000,F)& 0.195 &  -0.317 & 0.128 & 0.005 \\
\hline
\end{tabular}
\end{center}
\end{table}
}

In Section~\ref{sec:simulation}, we presented the bootstrapping result of correctly specified models. Here we have the bootstrapping result of Model 1 and Model 2 (See Table~\ref{tab:add1} and Table~\ref{tab:add2}, respectively) for settings including the result of both MAR and IPW, and for correctly specified model (T) and misspecified model (F).
\begin{table}[h]
\caption{Bootstrap confidence intervals for Model 1. \\
(True value of $\beta=1.8$ for both T and F settings).}
\label{tab:add1}
\begin{center}
\begin{tabular}{l|lllll}
\hline
\textbf{Setting} & \textbf{Est.} & \textbf{95\% CI} & \textbf{Width} & \textbf{Bias} \\
\hline %\\
IPW(n=500,T) & 1.941 & [1.594, 2.275] & 0.681 & 0.141 \\
IPW(n=1000,T) & 1.744 & [1.508, 1.970] & 0.462 & -0.056 \\
IPW(n=2000,T) & 1.767 & [1.598, 1.930] & 0.332 & -0.033 \\
MAR(n=500,T) & 2.350 & [2.052, 2.636] & 0.584 & 0.550 \\
MAR(n=1000,T) & 1.957 & [1.743, 2.184] & 0.441 & 0.157 \\
MAR(n=2000,T) & 2.007 & [1.869, 2.139] & 0.269 & 0.207 \\
IPW(n=500,F) & 1.630 & [1.279, 2.012] & 0.733 & -0.170 \\
IPW(n=1000,F) & 1.842 & [1.510, 2.136] & 0.627 & 0.042 \\
IPW(n=2000,F) & 1.727 & [1.442, 2.011] & 0.569 & -0.073 \\
MAR(n=500,F) & 1.337 & [1.013, 1.662] & 0.649 & -0.463 \\
MAR(n=1000,F) & 1.331 & [1.095, 1.580] & 0.484 & -0.469 \\
MAR(n=2000,F) & 1.213 & [1.053, 1.368] & 0.315 & -0.587 \\
\hline
\end{tabular}
\end{center}
\end{table}

\begin{table}[h]
\caption{Bootstrap confidence intervals for Model 2.
(True value of $\beta=-0.659$ for T and $\beta=-0.615$ for F settings).} 
\label{tab:add2}
\begin{center}
\begin{tabular}{l|lllll}
\hline
\textbf{Setting}  &\textbf{Est.} & \textbf{95\% CI} &    \textbf{Width} & \textbf{Bias} \\
\hline %\\
IPW(n=500,T) &  -0.557 &[-0.815, -0.303]& 0.513 & 0.102 \\
IPW(n=1000,T)& -0.760 &[-0.949, -0.569]& 0.380 & -0.101 \\
IPW(n=2000,T)& -0.707 &[-0.841, -0.561]& 0.281 & -0.048 \\
MAR(n=500,T) & -0.309 &[-0.571, -0.044]& 0.527 & 0.350 \\
MAR(n=1000,T)& -0.380 &[-0.585, -0.177]& 0.409 & 0.279 \\
MAR(n=2000,T)& -0.295 &[-0.447, -0.150]& 0.297 & 0.364 \\
IPW(n=500,F) & -0.502 &[-0.756, -0.275]& 0.481 & 0.113 \\
IPW(n=1000,F)& -0.621 &[-0.806, -0.435]& 0.371 & -0.006 \\
IPW(n=2000,F)& -0.625 &[-0.747, -0.505]& 0.242 & -0.010 \\
MAR(n=500,F) & -0.449 &[-0.739, -0.153]& 0.585 & 0.166 \\
MAR(n=1000,F)& -0.543 &[-0.738, -0.354]& 0.384 & 0.072 \\
MAR(n=2000,F)& -0.512 &[-0.644, -0.368]& 0.276 & 0.103 \\
\hline
\end{tabular}
\end{center}
\end{table}
\subsection{Application to COVID-19 case data}
\label{sec:app_covid}
In this study, we conducted meticulous data preprocessing on the COVID-19 Case Surveillance Public Use Data obtained from the Centers for Disease Control and Prevention (CDC), specifically focusing on New York state. 

We selected two time periods, March 2020 as the primary MNAR domain, and March 2023 as the auxillary MAR  conditions. The original data in March 2020 comprised $129,795$ records, while the  March 2023 subset contained $43,210$ records. The same preprocessing steps were applied to both datasets to filter out the low-quality entries: we excluded rows with missing values in critical columns such as the county, age group, sex, case-positive specimen interval, and exposure status. Additionally, cases were filtered to include only those with a "Laboratory-confirmed" status. Categorical variables were transformed into numerical indices, and a county score was recalculated into $[-1,1]$ to capture the relative positioning of each county. After these steps, the sample size of datasets in March 2020 and March 2023 are $125,737$ and $38,237$ respectively. Since we assumed the missingness mechanism is shared by race and hospitalization in primary domain, we further filtered the dataset in March 2020 accordingly and the final dataset consisted of $78,119$ patients.

\begin{table}[ht]
\caption{Bootstrap confidence intervals for application to COVID-19 dataset} 
\label{tab:add_covid}
\begin{center}
\begin{tabular}{l|lllll}
\hline
\textbf{n}  &\textbf{Est.} & \textbf{95\% CI} &    \textbf{Width}  \\
\hline %\\
Model 1 &  0.7396	&[0.7190,0.7570] & 0.0380	\\
Model 2& 0.7836 &[0.7558, 0.8071]& 0.0513 \\
MAR & 0.7337 &[0.7277,	0.7395]& 0.0117 \\
MCAR &  0.7533 & [0.7477, 0.7590] & 0.0112 \\
\hline
\end{tabular}
\end{center}
\end{table}

\end{document}